\documentclass[a4paper,11pt]{article}
\pdfoutput=1 
\usepackage{jheppub} 
\usepackage[T1]{fontenc} 
 \usepackage{verbatim}
\def\be{\begin{equation}}
\def\ee{\end{equation}}
\def\ba{\begin{eqnarray}}\def\bea{\begin{eqnarray}}
\def\ea{\end{eqnarray}}  \def\eea{\end{eqnarray}}

\def\nn{\nonumber}

\def\l{\lambda}
\def\O{\Omega}\def\o{\omega}
\def\th{\theta}

\def\[{\left[}
\def\]{\right]}
\def\({\left(}
\def\){\right)}

\def\<{\langle}
\def\>{\rangle}

   \def\sp2n{Sp(2N)}

\def\2F1{\,_2{\rm F}_1}

\def\tb{\tilde{b}}

\title{\boldmath No black hole bomb for D-dimensional extremal Reissner-Nordstrom black holes under charged massive scalar perturbation}
\author[a,b,c]{Jia-Hui Huang}

\affiliation[a]{Guangdong Provincial Key Laboratory of Quantum Engineering and Quantum Materials,
School of Physics and Telecommunication Engineering,South China Normal University,\\West Waihuan Road No.378, Higher Education Mega Center, Guangzhou, China}
\affiliation[b]{Guangdong Provincial Key Laboratory of Nuclear Science, Institute of quantum matter,South China Normal University,\\West Waihuan Road No.378, Higher Education Mega Center, Guangzhou, China}
\affiliation[c]{Guangdong-Hong Kong Joint Laboratory of Quantum Matter, South China Normal University,\\West Waihuan Road No.378, Higher Education Mega Center, Guangzhou, China}

\emailAdd{huangjh@m.scnu.edu.cn}

\abstract{The superradiant stability of asymptotically flat D-dimensional extremal Reissner-Nordstrom black holes under charged massive scalar perturbation is analytically studied. Recently, an analytical method has been proposed by the author and used to prove that five and six-dimensional extremal Reissner-Nordstrom black holes are superradiantly stable under charged massive scalar perturbation. We apply this analytical method  in the D-dimensional extremal Reissner-Nordstrom black hole cases and prove that the D-dimensional Reissner-Nordstrom black holes are all superradiantly stable under charged massive scalar perturbation. Our result is consistent with the previous numerical observation in the literature and provides a rigorous analytical proof.}

\begin{document}
\maketitle
\flushbottom

\section{Introduction}
Analysis of linear perturbation of black hole spacetimes plays an important role in many topics, such as the (in)stability of black hole solutions, the black hole ringdown phase after binary merger and astrophysics \cite{Regge:1957td,Barack:2018yly,Pani:2013pma}. Among various liner perturbation modes of black holes, superradiance mode is an interesting one, which can extract energy from the black holes\cite{Cardoso2004,Brito:2015oca,Brito:2014wla}.
 When a charged bosonic wave is scattering off a charged rotating black hole, the wave is amplified by the black hole if the angular frequency $\omega$ of the wave satisfies
  \begin{equation}\label{superRe}
   \omega < \text{m}\Omega_H  + e\Phi_H,
  \end{equation}
where $e$ and $\text{m}$ are the charge and azimuthal number of the bosonic wave mode, $\Omega_H$ is the angular velocity of the black hole horizon and $\Phi_H$ is the electromagnetic potential of the black hole horizon. This superradiant scattering was studied long time ago \cite{P1969,Ch1970,M1972,Ya1971,Bardeen1972,Bekenstein1973,Damour:1976kh}, and has broad applications in various areas of physics(for a recent comprehensive review, see\cite{Brito:2015oca}).

For a superradiant black-hole-and-perturbation system, if a mirror-like mechanism is introduced between the black hole horizon and spatial infinity, the amplified perturbation will be scattered back and forth between the "mirror" and black hole horizon, and this will lead to the superradiant instability of the system. This is dubbed black hole bomb mechanism \cite{PTbomb,Cardoso:2004nk,Herdeiro:2013pia,Degollado:2013bha}.
Superradiant (in)stability of various charged and rotating black holes has been studied extensively in the literature.
The superradiant (in)stability of four-dimensional rotating Kerr black holes under massive scalar or vector perturbation has been studied in \cite{Huang:2019xbu,Strafuss:2004qc,Konoplya:2006br,Cardoso:2011xi,Dolan:2012yt,Hod:2012zza,Hod:2014pza,Aliev:2014aba,Hod:2016iri,Degollado:2018ypf,Ponglertsakul:2020ufm,
East:2017ovw,East:2017mrj,Lin:2021ssw,Xu:2020fgq}.
Rotating or charged black holes with certain asymptotically curved space are proved to be superradiantly unstable under massless or massive bosonic perturbation \cite{Cardoso:2004hs,Cardoso:2013pza,Zhang:2014kna,Delice:2015zga,Aliev:2015wla,Wang:2015fgp,
Ferreira:2017tnc,Wang:2014eha,Bosch:2016vcp,Huang:2016zoz,Gonzalez:2017shu,Zhu:2014sya}, where the asymptotically curved geometries provide  natural mirror-like boundary conditions.

For asymptotically flat black holes, the  four-dimensional extremal and non-extremal Reissner-Nordstrom(RN) black holes have been proved superradiantly stable against charged massive scalar perturbation in the full parameter space of the black-hole-scalar-perturbation system\cite{Hod:2013eea,Huang:2015jza,Hod:2015hza,DiMenza:2014vpa,Mai:2021yny,Zou:2021mwa}. The argument in the proof is that the two conditions for the possible superradiant instability of the system, (i) existence of a trapping potential well outside the black hole horizon and (ii) superradiant amplification of the trapped modes, cannot be satisfied simultaneously in the RN-black-hole-scalar-perturbation system\cite{Hod:2013eea,Hod:2015hza}.

For various higher dimensional black holes, the linear stability analysis has also been studied in the literature (for an incomplete list, see\cite{Konoplya:2011qq,Konoplya:2007jv,Konoplya:2008au,Konoplya:2013sba,Konoplya:2008rq,Kodama:2003kk,Kodama:2007sf,Ishibashi:2011ws,Ishihara:2008re,Ishibashi:2003ap,Destounis:2019hca}).
In Ref.\cite{Konoplya:2008au}, the asymptotically flat RN black holes in D=5,6,..,11 are shown to be stable by studying the time-domain evolution of
the scalar perturbation with a numerical characteristic integration method. In Ref.\cite{Konoplya:2013sba}, the authors has provided numerical evidence that asymptotically flat extremal RN black holes are stable for arbitrary D.

Recently, an analytical method based on the \textit{Descartes' rule of signs} has been developed by the author to study the superradiant stability of higher dimensional RN black holes \cite{Huang:2021dpa,Huang:2021jaz}. Explicitly, it is proved that five and six dimensional extremal RN black holes and five dimensional non-extremal RN black holes are all superradiantly stable under charged massive scalar perturbation. These results are consistent with the previous analytical and numerical results.

In this work, we will go a step further and apply the above mentioned analytical method to study the superradiant stability of arbitrary D-dimensional (D$\geq 7$) extremal RN black hole under charged massive scalar perturbation. The effect on the dynamics of the scalar perturbation, which originates from the curved RN black holes, can be described an effective potential. We will show that there is no potential well for the effective potential experienced by the scalar perturbation. The two conditions for the possible superradiant instability can not be satisfied simultaneously, so D-dimensional extremal RN black hole are superradiantly stable under charged massive scalar perturbation.

The organization of this paper is as follows: In Section 2, we present a general description of the model and asymptotic analysis of boundary conditions. In Section 3, the effective potential of the radial equation of motion is given and the asymptotic behaviors of the effective potential at the horizon and spatial infinity are discussed.
In Section 4, it is shown that there is no potential well outside the black hole horizon for the superradiant modes. The final Section is devoted to the summary.

\section{Scalar field in D-dimensional RN black holes}
We first present our model with non-extremal D-dimensional RN black hole and then  take the extremal limit for further discussion.
The metric of the D-dimensional non-extremal RN black hole \cite{Myers:1986un,Huang:2021jaz,Huang:2021dpa} is
\bea
ds^2=-f(r)dt^2+\frac{dr^2}{f(r)}+r^2d\O_{D-2}^2.
\eea
The function $f(r)$ reads
\bea
f(r)=1-\frac{2m}{r^{D-3}}+\frac{q^2}{r^{2(D-3)}},
\eea
where the parameters $m$ and $q$ are related with the ADM mass $M$ and electric charge $Q$ of the RN black hole,
\bea
m=\frac{8\pi}{(D-2)Vol(S^{D-2})}M,~~ q= \frac{8\pi}{\sqrt{2(D-2)(D-3)}Vol(S^{D-2})}Q.
\eea
Here $Vol(S^{D-2})=2\pi^{\frac{D-1}{2}}/\Gamma(\frac{D-1}{2})$ is the volume of unit (D-2)-sphere.
$d\O_{D-2}^2$ is the common line element of a (D-2)-dimensional unit sphere $ S^{D-2}$ and can be written as
\bea
d\O_{D-2}^2=d\th_{D-2}^2+\sum^{D-3}_{i=1} \prod_{j=i+1}^{D-2}\sin^2(\th_{j})d\th_i^2,
\eea
 where the ranges of the angular coordinates are taken as $\th_i\in [0,\pi](i=2,..,D-2), \th_1\in [0,2\pi]$.
The inner and outer horizons of this RN black hole  are
  \bea
  r_\pm=(m\pm\sqrt{m^2-q^2})^{1/(D-3)}.
  \eea
It is obvious that we have the following two equalities
\bea
r_+^{D-3}+r_-^{D-3}=2m,~r_+^{D-3}r_-^{D-3}=q^2.
\eea
 The electromagnetic field outside the black hole horizon is described
 by the following 1-form  vector potential
 \bea
 A=-\sqrt{\frac{D-2}{2(D-3)}}\frac{q}{r^{D-3}} dt=-c_D\frac{q}{r^{D-3}} dt.
 \eea

The equation of motion for a charged massive scalar perturbation in this D-dimensional non-extremal black hole background is governed by the covariant Klein-Gordon equantion
\bea
(D_\nu D^\nu-\mu^2)\phi=0,
\eea
where $D_\nu=\nabla_\nu-ie A_\nu$ is the covariant derivative and $\mu,~e$ are the mass and charge of the scalar field respectively.
The solution of this equation with definite angular frequency  can be decomposed as
\bea
\phi(t,r,\th_i)=e^{-i\o t}R(r)\Theta(\th_i).
\eea
The angular eigenfunctions $\Theta(\th_i)$  are (D-2)-dimensional scalar spherical harmonics and the corresponding eigenvalues are given by $-l(l+D-3), (l=0,1,2,..)$\cite{Chodos:1983zi,Higuchi:1986wu,Rubin1984,Achour:2015zpa,Lindblom:2017maa}.

The radial equation of motion is described by
\bea\label{eq-radial}
\Delta\frac{d}{dr}(\Delta\frac{d R}{dr})+U R=0,
\eea
where
\bea\nn
\Delta&=&r^{D-2}f(r),\\
U&=&(\o+e A_t)^2 r^{2(D-2)}-l(l+D-3) r^{D-4}\Delta-\mu^2 r^{D-2}\Delta.
\eea

In order to analyze the physical boundary conditions needed here at the horizon and spatial infinity, we define the tortoise coordinate $y$ by $dy=\frac{r^{D-2}}{\Delta}dr$ and a new radial function $\tilde{R}=r^{\frac{D-2}{2}}R$, then the radial equation \eqref{eq-radial} can be rewritten as
\bea
\frac{d^2\tilde{R}}{dy^2}+\tilde{U} \tilde{R}=0,
\eea
where
\bea
\tilde{U}=\frac{U}{r^{2(D-2)}}-\frac{(D-2)f(r)[(D-4)f(r)+ 2 r f'(r)]}{4r^2}
\eea
The asymptotic behaviors of $\tilde{U}$ at the spatial infinity and the outer horizon are
\bea
\lim_{r\rightarrow +\infty}\tilde{U}= \o^2-\mu^2,~~
\lim_{r\rightarrow r_+} \tilde{U}= (\o-c_D\frac{e q}{r_+^{D-3}})^2=(\o-e\Phi_h)^2,
\eea
where $\Phi_h$ is the electric potential of at the outer horizon of the RN black hole.
Here we need purely ingoing wave condition at the horizon and bound state condition at spatial infinity, which leads to the following two conditions
\bea\label{sup-con}
\o&<&e\Phi_h=c_D\frac{e q}{r_+^{D-3}},\\
\label{bound-con}
\o&<&\mu.
\eea
The first inequality is the superradiance condition and the second inequality gives the bound sate condition.

\section{Effective potential and its asymptotic behaviors}
In order to analyze the superradiant stability of the RN black hole and scalar perturbation system, we define a new radial function $\psi=\Delta^{1/2} R$, then the radial equation of motion \eqref{eq-radial} can be written as a  Schrodinger-like equation
\bea
\frac{d^2\psi}{dr^2}+(\o^2-V)\psi=0,
\eea
where $V$ is the effective potential, which is the main object we will discuss. The explicit expression for the effective potential  $V$ is $V=\o^2+\frac{B_1}{A_1}$ and
\bea\label{V}
A_1&=&4r^{2}(r^{2D-6}-2 m r^{D-3}+q^2)^2,\\
B_1&=&4(\mu^2-\o^2)r^{4D-10}+(2l+D-2)(2l+D-4)r^{4D-12}-8(m\mu^2-c_D eq\o)r^{3D-7}\nn\\
&-&4m(2\l_l+(D-4)(D-2))r^{3D-9}+4q^2(\mu^2-c_D^2 e^2)r^{2D-4}\nn\\
&-&2(2m^2-q^2(2 \l_l+3(D-4)(D-2)+2))r^{2D-6}-4m q^2(D-4)(D-2)r^{D-3}\nn\\
&+&q^4(D-4)(D-2),
\eea
where $\l_l=l(l+D-3)$.

\subsection{Extremal limit}
Now  we consider the extremal limit by taking $m=q$.
In this limit, the superradiance condition becomes into
\bea\label{sup-con-extr}
\o<c_D e.
\eea
The expression of the effective potential $V$ becomes into
\bea\label{eff-pot}
V=\o^2+\frac{B}{A},
\eea
where $A$ and $B$ reads
\bea
A&=&4r^{2}(r^{2D-6}-2 m r^{D-3}+m^2)^2=4r^2(r^{D-3}-m)^4,\\
B&=&4(\mu^2-\o^2)r^{4D-10}+(2l+D-2)(2l+D-4)r^{4D-12}-8(m\mu^2-c_D e m \o)r^{3D-7}\nn\\
&-&4m(2\l_l+(D-4)(D-2))r^{3D-9}+4m^2(\mu^2-c_D^2 e^2)r^{2D-4}\nn\\
&+&2m^2(2 \l_l+3(D-4)(D-2))r^{2D-6}-4m^3(D-4)(D-2)r^{D-3}\nn\\
&+&m^4(D-4)(D-2)
\eea

In the extremal limit, the asymptotic behaviors of $V$  at the horizon and spatial infinity are
\bea
r\rightarrow r_h,~~ V\rightarrow -\infty;\\
r\rightarrow +\infty,~~ V\rightarrow \mu^2;
\eea
 At the spatial infinity, the asymptotic behavior of the derivative of the effective potential, $V'(r)$, is
 \bea\label{asymp}
V'(r)\rightarrow \left\{
   \begin{array}{ll}
     \frac{-(D-2)(D-4)-4\l_l-8m(\mu^2+c_D e \o-2\o^2)}{2r^3}, & \hbox{$D=5$;} \\
     \frac{-(D-2)(D-4)-4\l_l}{2r^3}, & \hbox{$ D\geqslant 6$.}
   \end{array}
 \right.
 \eea
Given the superradiance condition \eqref{sup-con-extr} and bound state condition \eqref{bound-con}, we can prove $V'(r)<0$ at spatial infinity when D=5. It is also obvious that $V'(r)<0$ at spatial infinity when D$\geqslant 6$. This means that there is no potential well near the spatial infinity and \textit{one maximum} exists for the
effective potential $V(r)$ outside the black hole horizon.

In the next section, we will prove that there is only one extreme (it is just the maximum discussed above) outside the event horizon $r_h$ for the effective potential in the D-dimensional extremal RN black hole case,  no potential well exists outside the event horizon for the superradiance modes. So there is no black hole bomb and D-dimensional extremal RN black holes are superradiantly stable under charged massive scalar perturbation. In our proof, the mathematical theorem\textit{ Descartes' rule of signs} plays an important role, which asserts that the number of positive roots of a polynomial equation with real coefficients is at most the number of sign changes in the sequence of the polynomial's coefficients.

\section{Analysis of the real roots of $V'=0$}
\label{root}
In this section, we show that there is only one extreme for the effective potential outside the RN black hole horizon by analyzing the derivative of the effective potential $V'(r)$. Explicitly, it is shown that only one real root exists for the following equation
\bea
V'(r)=0,
\eea
when $r>r_h$.

In the extremal RN black hole case, the derivative of effective potential \eqref{eff-pot} can be expressed as $V'(r)=\frac{E(r)}{F(r)}$.
$E(r)$ and $F(r)$ are polynomials of $r$, which read as follows
\bea
F(r)&=&2 r^3(r^{D-3}-m)^5,\\
E(r)&=&- (D_1 + 4 \l_l)r^{5(D-3)}+m(5D_1+(24-4D)\l_l)r^{4(D-3)}\nn\\
&&-2m^2(5D_1+(18-4D)\l_l)r^{3(D-3)}+2m^3(5D_1+(8-2D)\l_l)r^{2(D-3)}\nn\\
&&-4(D-3)m(\mu^2+\o(c_D e-2\o))r^{4D-10}+4(D-3)m^2(c_D^2 e^2+2\mu^2-3c_D e \o)r^{3D-7}\nn\\
&&+4(D-3)m^3(c_D^2 e^2-\mu^2)r^{2D-4}-5m^4D_1r^{D-3}+m^5D_1\nn\\
&=&a_0+a_1 r^{D-3}+a_2 r^{2D-6}+a'_2 r^{2D-4}+a_3 r^{3D-9}+a'_3 r^{3D-7}+a_4 r^{4D-12}+a'_4 r^{4D-10}+a_5 r^{5D-15},\nn\\
\eea
where
\bea\label{D-Er-xishu}
&&a_5=- (D_1 + 4 \l_l),~a_4=m(5D_1+(24-4D)\l_l),~a_3=-2m^2(5D_1+(18-4D)\l_l),\nn\\
&&a_2=2m^3(5D_1+(8-2D)\l_l),~a_1=-5m^4D_1,~a_0=m^5D_1,a'_4=-4(D-3)m(\mu^2+\o(c_D e-2\o))\nn\\
&&a'_3=4(D-3)m^2(c_D^2 e^2+2\mu^2-3c_D e \o),~a'_2=4(D-3)m^3(c_D^2 e^2-\mu^2).
\eea
and $D_1=D^2-6D+8=(D-2)(D-4)$.

Because we are interested in the real roots of the equation $V'(r)=0$, only the numerator $E(r)$ of $V'(r)$ is important for our analysis. It is equivalent to consider the real roots of the equation $E(r)=0$.
Making a change of variables from $r$ to $z=r-r_h$. $E(r)$ can be rewritten as a polynomial of $z$, $E(z)$. A real root of $E(r)=0$ when $r>r_h$ is equivalent to a positive root of $E(z)=0$.

In the following of this section, we will prove that there is only one positive real root for the equation $E(z)=0$, i.e., only one maximum  outside the event horizon $r_h$ for the effective potential in the D-dimensional extremal RN black hole case and no potential well exists outside the event horizon for the superradiance modes. This is achieved by
showing that for the sequence of the real coefficients $(b_{5D-15},b_{5D-16},...,b_0)$ in the polynomial $E(z)$, the sign change is always 1 and according to \textit{Descartes' rule of signs}, the equation $E(z)=0$  has at most one positive real root.

The polynomial of $E(z)$ can be expanded as
\bea
E(z)=\sum_{i=0}^{5D-15} b_i z^i.
\eea
The constant term $b_0$ in $E(z)$ is
\bea
b_0&=&a_0+a_1 r_h^{D-3}+a_2 r_h^{2D-6}+a'_2 r_h^{2D-4}+a_3 r_h^{3D-9}\nn\\&&+a'_3 r_h^{3D-7}+a_4 r_h^{4D-12}+a'_4 r_h^{4D-10}+a_5 r_h^{5D-15}.
\eea
Plugging \eqref{D-Er-xishu} into the above equation and after a straightforward calculation, we can obtain
\bea\label{f-0}
b_0=8(D-3)m^5 r_h^{2}(\o-c_D e)^2>0.
\eea
where we use the equation $r_h^{D-3}=m$.

It is easy to see that $a_5=- (D_1 + 4 \l_l)<0$. After considering the superradiance condition \eqref{sup-con-extr} and bound state condition \eqref{bound-con}, it is also easy to verify that
\bea
a'_4&=&-4(D-3)m(\mu^2+\o(c_D e-2\o))\nn\\
&=&-4(D-3)m(\mu^2-\o^2+\o(c_D e-\o))<0.
\eea
So we can immediately find that
\bea\label{f-1}
b_{5D-15}&=&a_{5}=- (D_1 + 4 \l_l)<0,\nn\\
b_{5D-16}&=&a_5 C_{5D-15}^1 r_h <0,\nn\\
...,\nn\\
b_{4D-9}&=&a_5 C_{5D-15}^{4D-9} r_h^{D-6}<0,\nn\\
b_{4D-10}&=&a_5 C_{5D-15}^{4D-10}r_h^{D-5}+a'_4<0,\nn\\
b_{4D-11}&=&a_5 C_{5D-15}^{4D-11}r_h^{D-4}+a'_4C_{4D-10}^{4D-11}r_h<0.
\eea
Then let us consider the coefficient $b_{4D-12}$ of $z^{4D-12}$, which is
\bea
b_{4D-12}=a_5 C_{5D-15}^{4D-12}r_h^{D-3}+a'_4 C_{4D-10}^{4D-12}r_h^{2}+a_4.
\eea
The term involving $a'_4$ is negative. Now, we show that the sum of the left two terms, $a_5 C_{5D-15}^{4D-12}r_h^{D-3}+a_4$, is also negative. It is easy to check that when $D\geq 7$, $C_{5D-15}^{4D-12}>5D-15>15$. Then
\bea
a_5 C_{5D-15}^{4D-12}r_h^{D-3}+a_4&=&-m D_1 (C_{5D-15}^{4D-12}-5)-4m \l_l (C_{5D-15}^{4D-12}-6+D),\nn
\eea
The first and second terms on the right of the above equation are both negative, so we have
\bea\label{f-2}
b_{4D-12}<0.
\eea

\subsection{Coefficients of $z^p$, $3D-7<p<4D-12$}
The coefficient of $z^p$ when $3D-7<p<4D-12$ can be written as follows,
\bea
b_p&=&a_5 C_{5D-15}^{p}r_h^{5D-15-p}+a'_4 C_{4D-10}^{p}r_h^{4D-10-p}+a_4C_{4D-12}^p r_h^{4D-12-p}\nn\\
&=&r_h^{4D-12-p}(a_5 C_{5D-15}^{p}m+a'_4C_{4D-10}^{p}r_h^{2}+a_4 C_{4D-12}^p )
\eea
On the right of the above equation, the term involving $a'_4$ is negative because  $a'_4<0$ . Now, let's prove the sum of the left two terms is also negative in the
following and  we will neglect the positive factor $r_h^{4D-12-p}$ for simplicity,
\bea\label{a4a5-3d}
&&a_5 C_{5D-15}^{p}m+a_4 C_{4D-12}^p\nn\\
&=&m C_{5D-15}^{p}(- D_1-4\l_l)+m C_{4D-12}^p(5D_1+4(6-D)\l_l)\nn\\
&=&m D_1(-C_{5D-15}^{p}+5C_{4D-12}^p)-4m\l_l  (C_{5D-15}^{p}+(D-6) C_{4D-12}^p)\nn\\
&=&m D_1C_{4D-12}^p(-\frac{5D-15}{4D-12}\cdot\frac{5D-16}{4D-13}\cdots\frac{5D-15-p+1}{4D-12-p+1}+5)\nn\\
&&-4m\l_l  (C_{5D-15}^{p}+(D-6) C_{4D-12}^p).
\eea
The $\l_l$ term in the above equation is obviously negative when $D\geq 7$. Because
\bea\label{c4c5-1}
-\frac{5D-15}{4D-12}\cdot\frac{5D-16}{4D-13}\cdots\frac{5D-15-p+1}{4D-12-p+1}<-(\frac{5}{4})^p,
\eea
and $(\frac{5}{4})^p>5$ when $p>3D-7>11$, the $D_1$ term in \eqref{a4a5-3d} is also negative. So \eqref{a4a5-3d} is negative.

We finally obtain that
\bea\label{f-3}
b_p<0,~~3D-7<p<4D-12.
\eea

\subsection{Coefficients of $z^p$, $p=3D-7,3D-8,3D-9$}
The three coefficients of  $z^p$, $p=3D-7,3D-8,3D-9$,  are listed as following
\bea
b_{3D-7}&=& a_5 C_{5D-15}^{3D-7}r_h^{2D-8}+a'_4 C_{4D-10}^{3D-7}r_h^{D-3}+a_4C_{4D-12}^{3D-7}r_h^{D-5}+a'_3,\nn\\
b_{3D-8}&=& a_5 C_{5D-15}^{3D-8}r_h^{2D-7}+a'_4 C_{4D-10}^{3D-8}r_h^{D-2}+a_4C_{4D-12}^{3D-8}r_h^{D-4}+a'_3
C_{3D-7}^{3D-8}r_h,\nn\\
b_{3D-9}&=& a_5 C_{5D-15}^{3D-9}r_h^{2D-6}+a'_4 C_{4D-10}^{3D-9}r_h^{D-1}+a_4C_{4D-12}^{3D-9}r_h^{D-3}+a'_3
C_{3D-7}^{3D-9}r_h^2+a_3.
\eea
Plugging \eqref{D-Er-xishu} into the above equations and after a straightforward calculation, we can obtain
\bea
b_{3D-7}&=&r_h^{D-5}(a_5 m C_{5D-15}^{3D-7}+a_4C_{4D-12}^{3D-7})+a'_4 m C_{4D-10}^{3D-7}+a'_3\nn\\
&=&r_h^{D-5}(-D_1 m(C_{5D-15}^{3D-7}-5C_{4D-12}^{3D-7})-4m\l_l(C_{5D-15}^{3D-7}+(D-6)C_{4D-12}^{3D-7}))\nn\\
&+&4(D-3)m^2(2C_{3D-7}^{3D-7}+C_{4D-10}^{3D-7})(c_D e-\o)(\frac{C_{3D-7}^{3D-7} c_D e}{2C_{3D-7}^{3D-7}+C_{4D-10}^{3D-7}}-\o)\nn\\
&+&4(D-3)m^2(\mu^2-\o^2)(2C_{3D-7}^{3D-7}-C_{4D-10}^{3D-7}),
\eea
\bea
b_{3D-8}&=&r_h^{D-4}(a_5 m C_{5D-15}^{3D-8}+a_4C_{4D-12}^{3D-8})+r_h(a'_4 m C_{4D-10}^{3D-8}+a'_3C_{3D-7}^{3D-8})\nn\\
&=&r_h^{D-4}(-D_1 m(C_{5D-15}^{3D-8}-5C_{4D-12}^{3D-8})-4m\l_l(C_{5D-15}^{3D-8}+(D-6)C_{4D-12}^{3D-8}))\nn\\
&+&4(D-3)m^2r_h(2C_{3D-7}^{3D-8}+C_{4D-10}^{3D-8})(c_D e-\o)(\frac{ C_{3D-7}^{3D-8}c_D e}{2C_{3D-7}^{3D-8}+C_{4D-10}^{3D-8}}-\o)\nn\\
&+&4(D-3)m^2r_h(\mu^2-\o^2)(2C_{3D-7}^{3D-8}-C_{4D-10}^{3D-8}),
\eea
\bea
b_{3D-9}&=&r_h^{D-3}(a_5 m C_{5D-15}^{3D-9}+a_4C_{4D-12}^{3D-9})+r_h^2(a'_4 m C_{4D-10}^{3D-9}+a'_3C_{3D-7}^{3D-9})+a_3\nn\\
&=&r_h^{D-3}(-D_1 m(C_{5D-15}^{3D-9}-5C_{4D-12}^{3D-9}+10)-4m\l_l(C_{5D-15}^{3D-9}+(D-6)C_{4D-12}^{3D-9}+9-2D))\nn\\
&+&4(D-3)m^2r_h^2(2C_{3D-7}^{3D-9}+C_{4D-10}^{3D-9})(c_D e-\o)(\frac{ C_{3D-7}^{3D-9} c_D e}{2C_{3D-7}^{3D-9}+C_{4D-10}^{3D-9}}-\o)\nn\\
&+&4(D-3)m^2r_h^2(\mu^2-\o^2)(2C_{3D-7}^{3D-9}-C_{4D-10}^{3D-9}).
\eea
Here it is not easy to fix the signs of the three coefficients.
Now we define the following three normalized coefficients
\bea
\tilde{b}_{3D-7}&=&\frac{r_h^{3D-7}}{2C_{3D-7}^{3D-7}+C_{4D-10}^{3D-7}}b_{3D-7},\nn\\
\tilde{b}_{3D-8}&=&\frac{r_h^{3D-8}}{2C_{3D-7}^{3D-8}+C_{4D-10}^{3D-8}}b_{3D-8},\nn\\
\tilde{b}_{3D-9}&=&\frac{r_h^{3D-9}}{2C_{3D-7}^{3D-9}+C_{4D-10}^{3D-9}}b_{3D-9}.
\eea
It is worth emphasizing that all the normalization factors are positive and the signs of $b_*$ and $\tilde{b}_*$ are the same. In the next, we will show that these coefficients satisfy
\bea\nn
\tilde{b}_{3D-7}<\tilde{b}_{3D-8},~~\tilde{b}_{3D-8}<\tilde{b}_{3D-9}.
\eea
When $\tilde{b}_{3D-7}<\tilde{b}_{3D-8}$, the possible signs of these two coefficients are $(-,-)(-,+),(+,+)$, which can be denoted as
$
\text{sign}(\tilde{b}_{3D-7})\leqslant\text{sign}(\tilde{b}_{3D-8}).
$
Equivalently, we have
$
\text{sign}(b_{3D-7})\leqslant\text{sign}(b_{3D-8}).
$

\subsubsection{$\tilde{b}_{3D-7}<\tilde{b}_{3D-8}$}
In this subsection, we will prove the following inequality,
\bea\label{diff-1}
\tilde{b}_{3D-8}-\tilde{b}_{3D-7}>0.
\eea
The difference $\tilde{b}_{3D-8}-\tilde{b}_{3D-7}$ can be divided into four terms and we will consider term by term in the following.

First, let's see the $\mu^2-\o^2$ term in the above difference, which can be written as
\bea\label{mu-o-term}
4(D-3)m^2r_h^{3D-7}(\mu^2-\o^2)[\frac{2C_{3D-7}^{3D-8}-C_{4D-10}^{3D-8}}{2C_{3D-7}^{3D-8}+C_{4D-10}^{3D-8}}
-\frac{2C_{3D-7}^{3D-7}-C_{4D-10}^{3D-7}}{2C_{3D-7}^{3D-7}+C_{4D-10}^{3D-7}}].
\eea
Given the bound state condition, the factor outside the square bracket in the above is positive. The factor in the square bracket is equivalent to
\bea\label{mu-o-factor}
\frac{-2C_{4D-10}^{3D-8}}{2C_{3D-7}^{3D-8}+C_{4D-10}^{3D-8}}
+\frac{2C_{4D-10}^{3D-7}}{2C_{3D-7}^{3D-7}+C_{4D-10}^{3D-7}}.
\eea
With the following combination identity
\bea\label{comb-1}
C_n^m=\frac{m+1}{n-m}C_n^{m+1},
\eea
we have
\bea\label{ImEq-1}
\frac{C_{3D-7}^{3D-8}}{C_{4D-10}^{3D-8}}=(D-2)\frac{C_{3D-7}^{3D-7}}{C_{4D-10}^{3D-7}}.
\eea
Then equation \eqref{mu-o-factor} can be rewritten as
\bea
\frac{-2}{2(D-2)C_{3D-7}^{3D-7}/C_{4D-10}^{3D-7}+1}
+\frac{2}{2C_{3D-7}^{3D-7}/C_{4D-10}^{3D-7}+1},
\eea
which is obviously positive and then equation \eqref{mu-o-term} is positive.

Second, let's see the $(c_D e-\o)$ term in the difference \eqref{diff-1},
\bea\label{e-o-1}
4(D-3)m^2r_h^{3D-7}(c_D e-\o)c_D e[\frac{C_{3D-7}^{3D-8}}{2C_{3D-7}^{3D-8}+C_{4D-10}^{3D-8}}
-\frac{C_{3D-7}^{3D-7}}{2C_{3D-7}^{3D-7}+C_{4D-10}^{3D-7}}].
\eea
Using the equation \eqref{ImEq-1}, the factor in the square bracket of the above can be rewritten as
\bea
\frac{(D-2)C_{3D-7}^{3D-7}}{2(D-2)C_{3D-7}^{3D-7}+C_{4D-10}^{3D-7}}
-\frac{C_{3D-7}^{3D-7}}{2C_{3D-7}^{3D-7}+C_{4D-10}^{3D-7}},
\eea
which is positive and then the equation \eqref{e-o-1} is positive.

Thirdly, let's see the $D_1$ term in the difference \eqref{diff-1},
\bea\label{D1-1}
D_1 m^5[-\frac{C_{5D-15}^{3D-8}-5C_{4D-12}^{3D-8}}{2C_{3D-7}^{3D-8}+C_{4D-10}^{3D-8}}+
\frac{C_{5D-15}^{3D-7}-5C_{4D-12}^{3D-7}}{2C_{3D-7}^{3D-7}+C_{4D-10}^{3D-7}}].
\eea
Using a similar proof as \eqref{c4c5-1}, we can obtain $C_{5D-15}^{3D-8}-5C_{4D-12}^{3D-8}>0,
C_{5D-15}^{3D-7}-5C_{4D-12}^{3D-7}>0$. In order to prove the factor in the square bracket in the above expression
is positive, we can equivalently to prove
\bea\nn
\frac{C_{5D-15}^{3D-8}-5C_{4D-12}^{3D-8}}{C_{5D-15}^{3D-7}-5C_{4D-12}^{3D-7}}&<&
\frac{2C_{3D-7}^{3D-8}+C_{4D-10}^{3D-8}}{2C_{3D-7}^{3D-7}+C_{4D-10}^{3D-7}}\\
\Leftrightarrow
\frac{\frac{1}{2D-7}C_{5D-15}^{3D-7}-5\frac{1}{D-4}C_{4D-12}^{3D-7}}{C_{5D-15}^{3D-7}-5C_{4D-12}^{3D-7}}&<&
\frac{2C_{3D-7}^{3D-7}+\frac{1}{D-2}C_{4D-10}^{3D-7}}{2C_{3D-7}^{3D-7}+C_{4D-10}^{3D-7}}.
\eea
Because
\bea
\frac{2C_{3D-7}^{3D-7}+\frac{1}{D-2}C_{4D-10}^{3D-7}}{2C_{3D-7}^{3D-7}+C_{4D-10}^{3D-7}}>\frac{1}{D-2}>
\frac{1}{2D-7}>
\frac{\frac{1}{2D-7}C_{5D-15}^{3D-7}-5\frac{1}{D-4}C_{4D-12}^{3D-7}}{C_{5D-15}^{3D-7}-5C_{4D-12}^{3D-7}},
\eea
we immediately obtain the expression \eqref{D1-1} is positive.

Finally, let's see the $\l_l$ term in the difference \eqref{diff-1},
\bea\label{l-1}
-4m^5\l_l[\frac{C_{5D-15}^{3D-8}+(D-6)C_{4D-12}^{3D-8}}{2C_{3D-7}^{3D-8}+C_{4D-10}^{3D-8}}
-\frac{C_{5D-15}^{3D-7}+(D-6)C_{4D-12}^{3D-7}}{2C_{3D-7}^{3D-7}+C_{4D-10}^{3D-7}}].
\eea
The positivity of the above expression is equivalent to
\bea
\frac{C_{5D-15}^{3D-8}+(D-6)C_{4D-12}^{3D-8}}{2C_{3D-7}^{3D-8}+C_{4D-10}^{3D-8}}
&<&\frac{C_{5D-15}^{3D-7}+(D-6)C_{4D-12}^{3D-7}}{2C_{3D-7}^{3D-7}+C_{4D-10}^{3D-7}}\\
\Leftrightarrow\frac{C_{5D-15}^{3D-8}+(D-6)C_{4D-12}^{3D-8}}{C_{5D-15}^{3D-7}+(D-6)C_{4D-12}^{3D-7}}
&<&\frac{2C_{3D-7}^{3D-8}+C_{4D-10}^{3D-8}}{2C_{3D-7}^{3D-7}+C_{4D-10}^{3D-7}}\\\label{diff-l-1}
\Leftrightarrow
\frac{\frac{1}{2D-7}C_{5D-15}^{3D-7}+(D-6)\frac{1}{D-4}C_{4D-12}^{3D-7}}{C_{5D-15}^{3D-7}+(D-6)C_{4D-12}^{3D-7}}
&<&\frac{2C_{3D-7}^{3D-7}+\frac{1}{D-2}C_{4D-10}^{3D-7}}{2C_{3D-7}^{3D-7}+C_{4D-10}^{3D-7}}.
\eea
The left side of the above inequality is
\bea
&&\frac{\frac{1}{2D-7}C_{5D-15}^{3D-7}+(D-6)\frac{1}{D-4}C_{4D-12}^{3D-7}}{C_{5D-15}^{3D-7}+(D-6)C_{4D-12}^{3D-7}}\nn\\
&=&\frac{\frac{1}{2D-7}k_D+\frac{1}{D-4}}{k_D+1}=\frac{1}{2D-7}+\frac{1}{k_D+1}(\frac{1}{D-4}-\frac{1}{2D-7}),
\eea
where
\bea
k_D=\frac{C_{5D-15}^{3D-7}}{(D-6)C_{4D-12}^{3D-7}}=\frac{1}{D-6}\cdot\frac{5D-15}{4D-12}\cdots
\frac{2D-7}{D-4}>\frac{1}{D-6}(\frac{5}{4})^{3D-7}>22.
\eea
So we have
\bea
\frac{1}{2D-7}+\frac{1}{k_D+1}(\frac{1}{D-4}-\frac{1}{2D-7})<
\frac{1}{2D-7}+\frac{1}{23}(\frac{1}{D-4}-\frac{1}{2D-7}).
\eea
Further, when $D\geqslant 7$ we have
\bea
\frac{1}{D-2}-(\frac{1}{2D-7}+\frac{1}{23}(\frac{1}{D-4}-\frac{1}{2D-7}))\nn\\
=\frac{1}{23(D-2)(D-4)(2D-7)}(22D^2-202D+454)>0.
\eea
The left side of the inequality \eqref{diff-l-1} is smaller than $\frac{1}{D-2}$.
One can easy to see that the right side of the inequality \eqref{diff-l-1} satisfies
\bea
\frac{2C_{3D-7}^{3D-7}+\frac{1}{D-2}C_{4D-10}^{3D-7}}{2C_{3D-7}^{3D-7}+C_{4D-10}^{3D-7}}>\frac{1}{D-2}.
\eea
The inequality \eqref{diff-l-1} holds when $D\geqslant 7$.

After showing that the four terms \eqref{mu-o-term} \eqref{e-o-1}\eqref{D1-1}\eqref{l-1} are all positive,
the inequality \eqref{diff-1} is proved when $D\geqslant 7$.
The possible signs for $b_{3D-7},b_{3D-8}$ are $(-,-),(-,+),(+,+)$.
The relation between signs of $b_{3D-7},b_{3D-8}$ is
\bea\label{f-4}
\text{sign}(b_{3D-7})\leqslant\text{sign}(b_{3D-8}).
\eea

\subsubsection{$\tilde{b}_{3D-8}<\tilde{b}_{3D-9}$}
In this subsection let's prove the following inequality
\bea\label{diff-2}
\tilde{b}_{3D-9}-\tilde{b}_{3D-8}>0.
\eea
The difference $\tilde{b}_{3D-9}-\tilde{b}_{3D-8}$ can be divided into four terms and we will consider term by term in the following.

First, let's see the $\mu^2-\o^2$ term in the above difference, which is
\bea\label{mu-o-2}
4(D-3)m^2 r_h^{3D-7}(\mu^2-\o^2)[\frac{2C_{3D-7}^{3D-9}-C_{4D-10}^{3D-9}}{2C_{3D-7}^{3D-9}+C_{4D-10}^{3D-9}}
-\frac{2C_{3D-7}^{3D-8}-C_{4D-10}^{3D-8}}{2C_{3D-7}^{3D-8}+C_{4D-10}^{3D-8}}]
\eea
The factor in the square bracket is equivalent to
\bea\label{mu-o-factor-2}
\frac{-2C_{4D-10}^{3D-9}}{2C_{3D-7}^{3D-9}+C_{4D-10}^{3D-9}}
+\frac{2C_{4D-10}^{3D-8}}{2C_{3D-7}^{3D-8}+C_{4D-10}^{3D-8}}\nn\\=
\frac{-2C_{4D-10}^{3D-8}}{(D-1)C_{3D-7}^{3D-8}+C_{4D-10}^{3D-8}}
+\frac{2C_{4D-10}^{3D-8}}{2C_{3D-7}^{3D-8}+C_{4D-10}^{3D-8}}.
\eea
Because $D\geqslant 7$, the above expression is positive. Thus the $\mu^2-\o^2$ term in the difference
$\tilde{b}_{3D-9}-\tilde{b}_{3D-8}$ is positive given the bound state condition.

Second, let's see the $(c_D e-\o)$ term in the difference $\tilde{b}_{3D-9}-\tilde{b}_{3D-8}$, which is
\bea\label{e-o-2}
4(D-3)m^2r_h^{3D-7}(c_D e-\o)c_D e[\frac{C_{3D-7}^{3D-9}}{2C_{3D-7}^{3D-9}+C_{4D-10}^{3D-9}}
-\frac{C_{3D-7}^{3D-8}}{2C_{3D-7}^{3D-8}+C_{4D-10}^{3D-8}}].
\eea
Define the factors in the square brackets in\eqref{mu-o-2} and \eqref{e-o-2}as following
\bea
x_2&=&\frac{2C_{3D-7}^{3D-9}-C_{4D-10}^{3D-9}}{2C_{3D-7}^{3D-9}+C_{4D-10}^{3D-9}}
-\frac{2C_{3D-7}^{3D-8}-C_{4D-10}^{3D-8}}{2C_{3D-7}^{3D-8}+C_{4D-10}^{3D-8}},\\
y_2&=&\frac{C_{3D-7}^{3D-9}}{2C_{3D-7}^{3D-9}+C_{4D-10}^{3D-9}}
-\frac{C_{3D-7}^{3D-8}}{2C_{3D-7}^{3D-8}+C_{4D-10}^{3D-8}}.
\eea
One can check that $2 y_2-\frac{x_2}{2}=0$. We have already proved $x_2>0$ in \eqref{mu-o-factor-2}, so
$y_2>0$, i.e. the $(c_D e-\o)$ term is positive.

Thirdly, let's see the $D_l$ term in the difference $\tilde{b}_{3D-9}-\tilde{b}_{3D-8}$, which is
\bea\label{D-2}
-D_1 m r_h^{4D-12}[\frac{C_{5D-15}^{3D-9}-5C_{4D-12}^{3D-9}+10}{2C_{3D-7}^{3D-9}+C_{4D-10}^{3D-9}}
-\frac{C_{5D-15}^{3D-8}-5C_{4D-12}^{3D-8}}{2C_{3D-7}^{3D-8}+C_{4D-10}^{3D-8}}].
\eea
Because
\bea\label{5d-4d}
\frac{C_{5D-15}^{3D-8}}{C_{4D-12}^{3D-8}}=\frac{5D-15}{4D-12}\frac{5D-16}{4D-13}\cdots\frac{2D-6}{D-3}>
(\frac{5}{4})^{3D-8}>(\frac{5}{4})^{10}\approx 9.3>5,
\eea
we have
\bea
C_{5D-15}^{3D-8}-5C_{4D-12}^{3D-8}>0.
\eea
Now let's prove the factor in square bracket is negative. The factor is
\bea\label{D-2-factor}
\frac{C_{5D-15}^{3D-9}-5C_{4D-12}^{3D-9}+10}{2C_{3D-7}^{3D-9}+C_{4D-10}^{3D-9}}
-\frac{C_{5D-15}^{3D-8}-5C_{4D-12}^{3D-8}}{2C_{3D-7}^{3D-8}+C_{4D-10}^{3D-8}}.
\eea
According to \eqref{comb-1}, the first term of the factor  can be written as
\bea\label{D-2-factor-1}
\frac{C_{5D-15}^{3D-9}-5C_{4D-12}^{3D-9}+10}{2C_{3D-7}^{3D-9}+C_{4D-10}^{3D-9}}
=\frac{\frac{3D-8}{2D-6}C_{5D-15}^{3D-8}-\frac{3D-8}{D-3}5C_{4D-12}^{3D-8}+10}
{\frac{3D-8}{2}2C_{3D-7}^{3D-8}+\frac{3D-8}{D-1}C_{4D-10}^{3D-8}}\nn\\
=\frac{C_{5D-15}^{3D-8}-2*5C_{4D-12}^{3D-8}+\frac{2D-6}{3D-8}*10}
{(D-3)2C_{3D-7}^{3D-8}+2\frac{D-3}{D-1}C_{4D-10}^{3D-8}}.
\eea
Then the negativity of \eqref{D-2-factor} is equivalent to
\bea\nn
\frac{C_{5D-15}^{3D-8}-2*5C_{4D-12}^{3D-8}+\frac{2D-6}{3D-8}*10}
{(D-3)2C_{3D-7}^{3D-8}+2\frac{D-3}{D-1}C_{4D-10}^{3D-8}}
<\frac{C_{5D-15}^{3D-8}-5C_{4D-12}^{3D-8}}{2C_{3D-7}^{3D-8}+C_{4D-10}^{3D-8}}\\
\label{D-2-factor-2}
\Leftrightarrow
\frac{C_{5D-15}^{3D-8}-2*5C_{4D-12}^{3D-8}+\frac{2D-6}{3D-8}*10}
{C_{5D-15}^{3D-8}-5C_{4D-12}^{3D-8}}
<\frac{(D-3)2C_{3D-7}^{3D-8}+2\frac{D-3}{D-1}C_{4D-10}^{3D-8}}{2C_{3D-7}^{3D-8}+C_{4D-10}^{3D-8}}.
\eea
For the left side of the equation \eqref{D-2-factor-2}, we have
\bea
\frac{C_{5D-15}^{3D-8}-2*5C_{4D-12}^{3D-8}+\frac{2D-6}{3D-8}*10}
{C_{5D-15}^{3D-8}-5C_{4D-12}^{3D-8}}=
1+5*\frac{-C_{4D-12}^{3D-8}+\frac{4D-12}{3D-8}}{C_{5D-15}^{3D-8}-5C_{4D-12}^{3D-8}}<1.
\eea
For the right side of the equation \eqref{D-2-factor-2}, we have
\bea
\frac{(D-3)2C_{3D-7}^{3D-8}+2\frac{D-3}{D-1}C_{4D-10}^{3D-8}}{2C_{3D-7}^{3D-8}+C_{4D-10}^{3D-8}}
=\frac{2D-6}{D-1}\frac{(D-1)C_{3D-7}^{3D-8}+C_{4D-10}^{3D-8}}{2C_{3D-7}^{3D-8}+C_{4D-10}^{3D-8}}
>\frac{2D-6}{D-1}.
\eea
When $D>6$, we have $\frac{2D-6}{D-1}>1$. Then the equation \eqref{D-2-factor} is negative and the $D_1$ term
,\eqref{D-2}, is positive.

Finally, let's see the $\l_l$ term in the difference $\tilde{b}_{3D-9}-\tilde{b}_{3D-8}$, which is
\bea\label{l-2}
-4m\l_l r_h^{4D-12}[\frac{C_{5D-15}^{3D-9}+(D-6)C_{4D-12}^{3D-9}+9-2D}{2C_{3D-7}^{3D-9}+C_{4D-10}^{3D-9}}
-\frac{C_{5D-15}^{3D-8}+(D-6)C_{4D-12}^{3D-8}}{2C_{3D-7}^{3D-8}+C_{4D-10}^{3D-8}}].
\eea
Now let's prove the factor in square bracket is negative. This factor is
\bea\label{l-2-factor}
\frac{C_{5D-15}^{3D-9}+(D-6)C_{4D-12}^{3D-9}+9-2D}{2C_{3D-7}^{3D-9}+C_{4D-10}^{3D-9}}
-\frac{C_{5D-15}^{3D-8}+(D-6)C_{4D-12}^{3D-8}}{2C_{3D-7}^{3D-8}+C_{4D-10}^{3D-8}}.
\eea
Since $9-2D<0$,  a sufficient condition for the above to be negative is
\bea\label{l-2-factor-1}
\frac{C_{5D-15}^{3D-9}+(D-6)C_{4D-12}^{3D-9}}{2C_{3D-7}^{3D-9}+C_{4D-10}^{3D-9}}
<\frac{C_{5D-15}^{3D-8}+(D-6)C_{4D-12}^{3D-8}}{2C_{3D-7}^{3D-8}+C_{4D-10}^{3D-8}}.
\eea
According to \eqref{comb-1}, the left side of the above can be rewritten as
\bea
\frac{C_{5D-15}^{3D-9}+(D-6)C_{4D-12}^{3D-9}}{2C_{3D-7}^{3D-9}+C_{4D-10}^{3D-9}}=
\frac{\frac{1}{2D-6}C_{5D-15}^{3D-8}+\frac{1}{D-3}(D-6)C_{4D-12}^{3D-8}}{\frac{1}{2}2C_{3D-7}^{3D-8}+\frac{1}{D-1}C_{4D-10}^{3D-8}}
\nn\\
=\frac{C_{5D-15}^{3D-8}+2(D-6)C_{4D-12}^{3D-8}}{(D-3)2C_{3D-7}^{3D-8}+\frac{2(D-3)}{D-1}C_{4D-10}^{3D-8}}.
\eea
The inequality \eqref{l-2-factor-1} is equivalent to
\bea\nn
\frac{C_{5D-15}^{3D-8}+2(D-6)C_{4D-12}^{3D-8}}{(D-3)2C_{3D-7}^{3D-8}+\frac{2(D-3)}{D-1}C_{4D-10}^{3D-8}}
<\frac{C_{5D-15}^{3D-8}+(D-6)C_{4D-12}^{3D-8}}{2C_{3D-7}^{3D-8}+C_{4D-10}^{3D-8}}\\ \nn
\Leftrightarrow
\frac{C_{5D-15}^{3D-8}+2(D-6)C_{4D-12}^{3D-8}}{C_{5D-15}^{3D-8}+(D-6)C_{4D-12}^{3D-8}}
<\frac{(D-3)2C_{3D-7}^{3D-8}+\frac{2(D-3)}{D-1}C_{4D-10}^{3D-8}}{2C_{3D-7}^{3D-8}+C_{4D-10}^{3D-8}}\\
\Leftrightarrow
1+\frac{D-6}{D-6+\frac{C_{5D-15}^{3D-8}}{C_{4D-12}^{3D-8}}}<\frac{2D-6}{D-1}+\frac{2(D-3)^2/(D-1)}{2+
\frac{C_{4D-10}^{3D-8}}{C_{3D-7}^{3D-8}}},
\eea
According to \eqref{5d-4d}, when $D>6$, we have
\bea
1+\frac{D-6}{D-6+\frac{C_{5D-15}^{3D-8}}{C_{4D-12}^{3D-8}}}<1+\frac{D-6}{D+3}=\frac{2D-3}{D+3}<
\frac{2D-6}{D-1}.
\eea
Thus, the factor \eqref{l-2-factor} is negative and the $\l_l$ term, Eq.\eqref{l-2}, is positive.

After proving that all the four terms \eqref{mu-o-2}\eqref{e-o-2}\eqref{D-2}\eqref{l-2} are positive, then
\eqref{diff-2} is proved when $D>6$. The relation between signs of $b_{3D-9}, b_{3D-8}$ is
\bea\label{f-5}
\text{sign}(b_{3D-8})\leqslant \text{sign}(b_{3D-9}).
\eea

\subsection{Coefficients of $z^p$, $ 3D-9> p>2D-4$}
Similarly, in this subsection, we will consider the sign relations between pairs of adjacent coefficients of $z^p$ for $ 3D-9> p>2D-4$. Explicitly,
we will analyze the signs of $b_p$ and $b_{p+1}$ by comparing $\tilde{b}_p$ and $\tilde{b}_{p+1}$. The expression of the coefficient $b_p$ is
\bea
b_p&=&a_5 C_{5D-15}^{p}r_h^{5D-15-p}+a'_4 C_{4D-10}^{p}r_h^{4D-10-p}\nn\\&&+a_4C_{4D-12}^{p}r_h^{4D-12-p}
+a'_3
C_{3D-7}^{p}r_h^{3D-7-p}+a_3 C_{3D-9}^p r_h^{3D-9-p}\nn\\
&=&r_h^{3D-9-p}(a_5 m^2 C_{5D-15}^{p}+a_4m C_{4D-12}^{p}+a_3 C_{3D-9}^p\nn\\&&~~~~~~~~~~
+a'_4 m r_h^2 C_{4D-10}^{p}+a'_3
r_h^2C_{3D-7}^{p})\nn\\
&=& r_h^{-p}(-m^5 D_1)(C_{5D-15}^{p}-5C_{4D-12}^{p}+10C_{3D-9}^p)\nn\\
&+&r_h^{-p}(-4m^5\l_l)(C_{5D-15}^{p}+(D-6)C_{4D-12}^{p}+(9-2D)C_{3D-9}^p)\nn\\
&+&r_h^{-p+2}4(D-3)m^5(C_{4D-10}^{p}+2 C_{3D-7}^p)(c_D e- \o)(\frac{ C_{3D-7}^p c_D e}{C_{4D-10}^{p}+2 C_{3D-7}^p}-\o)\nn\\
&+&r_h^{-p+2}4(D-3)m^5(\mu^2-\o^2)(2C_{3D-7}^p-C_{4D-10}^p).
\eea
Define a normalized coefficient $\tilde{b}_p$ with a positive factor,
\bea
\tb_p=\frac{r_h^p}{C_{4D-10}^{p}+2 C_{3D-7}^p}b_p.
\eea
Then we consider the following difference
\bea\label{diff-3}
\tb_p-\tb_{p+1}=\frac{r_h^p}{C_{4D-10}^{p}+2 C_{3D-7}^{p}}b_p-\frac{r_h^{p+1}}{C_{4D-10}^{p+1}+2 C_{3D-7}^{p+1}}b_{p+1},
\eea
which can be similarly decomposed into a sum of four terms.

First, the $(\mu^2-\o^2)$ term in the above difference is
\bea\label{mu-o-3}
4(D-3)m^5r_h^2(\mu^2-\o^2)[\frac{2 C_{3D-7}^p-C_{4D-10}^{p}}{C_{4D-10}^{p}+2 C_{3D-7}^p}-
\frac{2 C_{3D-7}^{p+1}-C_{4D-10}^{p+1}}{C_{4D-10}^{p+1}+2 C_{3D-7}^{p+1}}]
\eea
The factor in the square bracket is equivalent to $x_3$, which is defined as
\bea\label{mu-o-3-factor}
&&x_3\equiv\frac{-2C_{4D-10}^{p}}{C_{4D-10}^{p}+2 C_{3D-7}^p}+
\frac{2 C_{4D-10}^{p+1}}{C_{4D-10}^{p+1}+2 C_{3D-7}^{p+1}}\nn\\
&=&\frac{-2C_{4D-10}^{p+1}*\frac{1}{4D-10-p}}{\frac{1}{4D-10-p}C_{4D-10}^{p+1}+2 C_{3D-7}^{p+1}
*\frac{1}{3D-7-p}}+
\frac{2 C_{4D-10}^{p+1}}{C_{4D-10}^{p+1}+2 C_{3D-7}^{p+1}}\nn\\
&=&\frac{-2C_{4D-10}^{p+1}}{C_{4D-10}^{p+1}+2 C_{3D-7}^{p+1}
*\frac{4D-10-p}{3D-7-p}}+
\frac{2 C_{4D-10}^{p+1}}{C_{4D-10}^{p+1}+2 C_{3D-7}^{p+1}}.
\eea
Because $\frac{4D-10-p}{3D-7-p}>1$ , the above expression is positive. Then \eqref{mu-o-3} is positive given the bound state condition.

Secondly, the $(c_D e-\o)$ term in the difference \eqref{diff-3} is
\bea\label{e-o-3}
4(D-3)m^5r_h^2(c_D e-\o)c_D e[\frac{ C_{3D-7}^p}{C_{4D-10}^{p}+2 C_{3D-7}^p}-
\frac{C_{3D-7}^{p+1}}{C_{4D-10}^{p+1}+2 C_{3D-7}^{p+1}}].
\eea
The factor in the above square bracket can be denoted as
\bea
y_3\equiv\frac{ C_{3D-7}^p}{C_{4D-10}^{p}+2 C_{3D-7}^p}-
\frac{C_{3D-7}^{p+1}}{C_{4D-10}^{p+1}+2 C_{3D-7}^{p+1}}.
\eea
One can easily check that
\bea
2 y_3-x_3/2=0.
\eea
And because $x_3>0$, we then have $y_3>0$. Together with the superradiance condition, we obtain that the $(c_D e-\o)$ term is positive.

Thirdly, let's see the $\l_l$ term in the difference \eqref{diff-3}, which is
\bea\label{l-3}
-4m^5\l_l[\frac{C_{5D-15}^{p}+(D-6)C_{4D-12}^{p}+(9-2D)C_{3D-9}^p}{C_{4D-10}^{p}+2 C_{3D-7}^{p}}\nn\\-
\frac{C_{5D-15}^{p+1}+(D-6)C_{4D-12}^{p+1}+(9-2D)C_{3D-9}^{p+1}}{C_{4D-10}^{p+1}+2 C_{3D-7}^{p+1}}].
\eea
The factor in square bracket is
\bea\label{l-3-factor}
\frac{C_{5D-15}^{p}+(D-6)C_{4D-12}^{p}}{C_{4D-10}^{p}+2 C_{3D-7}^{p}}
-\frac{C_{5D-15}^{p+1}+(D-6)C_{4D-12}^{p+1}}{C_{4D-10}^{p+1}+2 C_{3D-7}^{p+1}}
\nn\\
+\frac{(2D-9)C_{3D-9}^{p+1}}{C_{4D-10}^{p+1}+2 C_{3D-7}^{p+1}}
-\frac{(2D-9)C_{3D-9}^p}{C_{4D-10}^{p}+2 C_{3D-7}^{p}}.
\eea
For the second line of the above expression, we have
\bea
&&\frac{(2D-9)C_{3D-9}^{p+1}}{C_{4D-10}^{p+1}+2 C_{3D-7}^{p+1}}
-\frac{(2D-9)C_{3D-9}^p}{C_{4D-10}^{p}+2 C_{3D-7}^{p}}\nn\\
&=&(2D-9)[\frac{(3D-9-p)C_{3D-9}^{p}}{(4D-10-p)C_{4D-10}^{p}+(3D-7-p)2 C_{3D-7}^{p}}
-\frac{C_{3D-9}^p}{C_{4D-10}^{p}+2 C_{3D-7}^{p}}]\nn\\
&=&(2D-9)[\frac{C_{3D-9}^{p}}{\frac{4D-10-p}{3D-9-p}C_{4D-10}^{p}+\frac{3D-7-p}{3D-9-p}2 C_{3D-7}^{p}}
-\frac{C_{3D-9}^p}{C_{4D-10}^{p}+2 C_{3D-7}^{p}}].
\eea
Since $\frac{4D-10-p}{3D-9-p}>1, \frac{3D-7-p}{3D-9-p}>1 $, the above equation is less than 0. So the
second line of \eqref{l-3-factor} is negative.

For the first line of \eqref{l-3-factor}, we will show that it is also negative, which is equivalent to
\bea
&&\frac{C_{5D-15}^{p}+(D-6)C_{4D-12}^{p}}{C_{4D-10}^{p}+2 C_{3D-7}^{p}}
<\frac{C_{5D-15}^{p+1}+(D-6)C_{4D-12}^{p+1}}{C_{4D-10}^{p+1}+2 C_{3D-7}^{p+1}}\nn\\
&\Leftrightarrow&
\frac{C_{4D-10}^{p+1}+2 C_{3D-7}^{p+1}}{C_{4D-10}^{p}+2 C_{3D-7}^{p}}
<\frac{C_{5D-15}^{p+1}+(D-6)C_{4D-12}^{p+1}}{C_{5D-15}^{p}+(D-6)C_{4D-12}^{p}}\nn\\
&\Leftrightarrow&
\frac{(4D-10-p)C_{4D-10}^{p}+(3D-7-p)2 C_{3D-7}^{p}}{C_{4D-10}^{p}+2 C_{3D-7}^{p}}\nn\\
\label{l-3-factor-1}
&&<\frac{(5D-15-p)C_{5D-15}^{p}+(4D-12-p)(D-6)C_{4D-12}^{p}}{C_{5D-15}^{p}+(D-6)C_{4D-12}^{p}}
\eea
For the left side of the above inequality, we have
\bea\label{l-3-factor-4}
\frac{(4D-10-p)C_{4D-10}^{p}+(3D-7-p)2 C_{3D-7}^{p}}{C_{4D-10}^{p}+2 C_{3D-7}^{p}}<4D-10-p.
\eea
For the right side of the inequality \eqref{l-3-factor-1}, we have
\bea\label{l-3-factor-2}
\frac{(5D-15-p)C_{5D-15}^{p}+(4D-12-p)(D-6)C_{4D-12}^{p}}{C_{5D-15}^{p}+(D-6)C_{4D-12}^{p}}\nn\\
=4D-12-p+\frac{D-3}{1+(D-6)C_{4D-12}^p/C_{5D-15}^p}.
\eea
Since
\bea
\frac{C_{4D-12}^p}{C_{5D-15}^p}=\frac{4D-12}{5D-15}\frac{4D-13}{5D-16}\cdots\frac{4D-12-p}{5D-15-p}
<(\frac{4}{5})^p<(\frac{4}{5})^{2D-4}<(\frac{4}{5})^{8}<0.2,
\eea
then for $D>6$
\bea\label{l-3-factor-3}
4D-12-p+\frac{D-3}{1+(D-6)C_{4D-12}^p/C_{5D-15}^p}
>4D-12-p+\frac{D-3}{1+(D-6)0.2}\nn\\
>4D-12-p+3=4D-9-p.
\eea
Based on \eqref{l-3-factor-4}, \eqref{l-3-factor-3}, we obtain that the first line of \eqref{l-3-factor} is also negative.  So the $\l_l$ term, \eqref{l-3}, is positive.

Finally, let's see the $D_1$ term in the difference \eqref{diff-3}, which is
\bea\label{D-3}
-m^5 D_1[\frac{C_{5D-15}^{p}-5C_{4D-12}^{p}+10C_{3D-9}^p}{C_{4D-10}^{p}+2 C_{3D-7}^p}-
\frac{C_{5D-15}^{p+1}-5C_{4D-12}^{p+1}+10C_{3D-9}^{p+1}}{C_{4D-10}^{p+1}+2 C_{3D-7}^{p+1}}].
\eea
We will prove the difference in the above square bracket is negative and then $D_1$ term is positive.

For $D>6$ and $p>2D-4>8$, we have the following inequalities
\bea
\frac{C_{5D-15}^{p}}{C_{3D-9}^{p}}=\frac{5D-15}{3D-9}\frac{5D-16}{3D-10}..\frac{5D-14-p}{3D-8-p}>(\frac{5}{3})^8>10,\\
\frac{C_{3D-9}^{p}}{C_{5D-15}^{p}}<(3/5)^8<0.02,\\
\frac{C_{5D-15}^{p}}{C_{4D-12}^{p}}=\frac{5D-15}{4D-12}\frac{5D-16}{4D-13}..\frac{5D-14-p}{4D-11-p}>(\frac{5}{4})^8>5,
\eea
so $C_{5D-15}^{p}-5C_{4D-12}^{p}+10C_{3D-9}^p>0$, i.e. the numerators in \eqref{D-3} are positive. Then
\bea
&&\frac{C_{5D-15}^{p}-5C_{4D-12}^{p}+10C_{3D-9}^p}{C_{4D-10}^{p}+2 C_{3D-7}^p}-
\frac{C_{5D-15}^{p+1}-5C_{4D-12}^{p+1}+10C_{3D-9}^{p+1}}{C_{4D-10}^{p+1}+2 C_{3D-7}^{p+1}}<0\nn\\
&\Leftrightarrow& \frac{C_{4D-10}^{p+1}+2 C_{3D-7}^{p+1}}{C_{4D-10}^{p}+2 C_{3D-7}^p}<
\frac{C_{5D-15}^{p+1}-5C_{4D-12}^{p+1}+10C_{3D-9}^{p+1}}{C_{5D-15}^{p}-5C_{4D-12}^{p}+10C_{3D-9}^p}\\ \label{D-3-coef}
&\Leftrightarrow& \frac{(4D-10-p)C_{4D-10}^{p}+(3D-7-p)2 C_{3D-7}^{p}}{C_{4D-10}^{p}+2 C_{3D-7}^p}<\nn\\
&&\frac{(5D-15-p)C_{5D-15}^{p}-(4D-12-p)5C_{4D-12}^{p}+(3D-9-p)10C_{3D-9}^{p}}{C_{5D-15}^{p}-5C_{4D-12}^{p}+10C_{3D-9}^p}.
\eea
When $D>6$, the left term in inequality \eqref{D-3-coef} satisfies
\bea\label{D-3-coef-left}
\frac{(4D-10-p)C_{4D-10}^{p}+(3D-7-p)2 C_{3D-7}^{p}}{C_{4D-10}^{p}+2 C_{3D-7}^p}<4D-10-p.
\eea
For $D>6$, we also have $D-3>3$. The right term in inequality \eqref{D-3-coef} satisfies
\bea
&&\frac{(5D-15-p)C_{5D-15}^{p}-(4D-12-p)5C_{4D-12}^{p}+(3D-9-p)10C_{3D-9}^{p}}{C_{5D-15}^{p}-5C_{4D-12}^{p}+10C_{3D-9}^p}\nn\\
&=&4D-12-p+\frac{(D-3)C_{5D-15}^{p}-(D-3)10C_{3D-9}^{p}}{C_{5D-15}^{p}-5C_{4D-12}^{p}+10C_{3D-9}^p}\nn\\
&>&4D-12-p+(D-3)\frac{C_{5D-15}^{p}-10C_{3D-9}^{p}}{C_{5D-15}^{p}+10C_{3D-9}^p}\nn\\
&=&4D-12-p+(D-3)\frac{1-10C_{3D-9}^{p}/C_{5D-15}^{p}}{1+10C_{3D-9}^p/C_{5D-15}^{p}}\nn\\
&>&4D-12-p+(D-3)\frac{1-10*0.02}{1+10*0.02}\nn\\
&>&4D-12-p+3*\frac{2}{3}=4D-10-p.
\eea
With the above inequality and \eqref{D-3-coef-left}, we obtain that the $D_1$ term \eqref{D-3} is positive.

So in this case, all four terms in the difference \eqref{diff-3} are shown to be positive, we then obtain that
\bea\label{f-6}
\text{sign}(b_p)\geqslant \text{sign}(b_{p+1}).
\eea

\subsection{Coefficients of $z^p$, $ p=2D-3, 2D-4$}
The two coefficients of $z^p$ for $ p=2D-3, 2D-4$ are listed as follows
\bea
b_{2D-3}&=&a_5 C_{5D-15}^{2D-3}r_h^{3D-12}+a'_4 C_{4D-10}^{2D-3}r_h^{2D-7}+a_4C_{4D-12}^{2D-3}r_h^{2D-9}
\nn\\&&+a'_3 C_{3D-7}^{2D-3}r_h^{D-4}+a_3 C_{3D-9}^{2D-3} r_h^{D-6}\nn\\
&=& r_h^{-3}(-m^3 D_1)(C_{5D-15}^{2D-3}-5C_{4D-12}^{2D-3}+10C_{3D-9}^{2D-3})\nn\\
&+&r_h^{-3}(-4m^3\l_l)(C_{5D-15}^{2D-3}+(D-6)C_{4D-12}^{2D-3}+(9-2D)C_{3D-9}^{2D-3})\nn\\
&+&r_h^{-1}4(D-3)m^3(C_{4D-10}^{2D-3}+2 C_{3D-7}^{2D-3})(c_D e- \o)(\frac{ C_{3D-7}^{2D-3} c_D e}{C_{4D-10}^{2D-3}+2 C_{3D-7}^{2D-3}}-\o)\nn\\
&+&r_h^{-1}4(D-3)m^3(\mu^2-\o^2)(2C_{3D-7}^{2D-3}-C_{4D-10}^{2D-3}).
\eea
\bea
b_{2D-4}&=&a_5 C_{5D-15}^{2D-4}r_h^{3D-11}+a'_4 C_{4D-10}^{2D-4}r_h^{2D-6}+a_4C_{4D-12}^{2D-4}r_h^{2D-8}
\nn\\&&+a'_3 C_{3D-7}^{2D-4}r_h^{D-3}+a_3 C_{3D-9}^{2D-4} r_h^{D-5}+a'_2 C_{2D-4}^{2D-4}\nn\\
&=& r_h^{-2}(-m^3 D_1)(C_{5D-15}^{2D-4}-5C_{4D-12}^{2D-4}+10C_{3D-9}^{2D-4})\nn\\
&+&r_h^{-2}(-4m^3\l_l)(C_{5D-15}^{2D-4}+(D-6)C_{4D-12}^{2D-4}+(9-2D)C_{3D-9}^{2D-4})\nn\\
&+&4(D-3)m^3(C_{4D-10}^{2D-4}+2 C_{3D-7}^{2D-4}-1)(c_D e- \o)(\frac{ (C_{3D-7}^{2D-4}+1) c_D e}{C_{4D-10}^{2D-4}+2 C_{3D-7}^{2D-4}-1}-\o)\nn\\
&+&4(D-3)m^3(\mu^2-\o^2)(2C_{3D-7}^{2D-4}-C_{4D-10}^{2D-4}-1).
\eea
Define two normalized new coefficients with positive factors,
\bea
\tb_{2D-3}=\frac{b_{2D-3}r_h^3}{C_{4D-10}^{2D-3}+2 C_{3D-7}^{2D-3}},~
\tb_{2D-4}=\frac{b_{2D-4}r_h^2}{C_{4D-10}^{2D-4}+2 C_{3D-7}^{2D-4}-1}.
\eea
Now, we consider the difference between $\tb_{2D-3}$ and $\tb_{2D-4}$
\bea\label{diff-4}
\tb_{2D-4}-\tb_{2D-3}=\frac{b_{2D-4}r_h^2}{C_{4D-10}^{2D-4}+2 C_{3D-7}^{2D-4}-1}
-\frac{b_{2D-3}r_h^3}{C_{4D-10}^{2D-3}+2 C_{3D-7}^{2D-3}}.
\eea
The difference can be decomposed into four terms and we will analyze term by term.

First, let's see the $(\mu^2-\o^2)$ term in \eqref{diff-4}, which is
\bea\label{mu-o-4}
4(D-3)m^3 r_h^2(\mu^2-\o^2)[\frac{2C_{3D-7}^{2D-4}-C_{4D-10}^{2D-4}-1}{C_{4D-10}^{2D-4}+2 C_{3D-7}^{2D-4}-1}
-\frac{2C_{3D-7}^{2D-3}-C_{4D-10}^{2D-3}}{C_{4D-10}^{2D-3}+2 C_{3D-7}^{2D-3}}]
\eea
The positivity of the factor in the square bracket is equivalent to
\bea
&&\frac{-2C_{4D-10}^{2D-4}}{C_{4D-10}^{2D-4}+2 C_{3D-7}^{2D-4}-1}
+\frac{2C_{4D-10}^{2D-3}}{C_{4D-10}^{2D-3}+2 C_{3D-7}^{2D-3}}>0\nn\\
&\Leftrightarrow&
\frac{C_{4D-10}^{2D-3}}{C_{4D-10}^{2D-3}+2 C_{3D-7}^{2D-3}}>\frac{C_{4D-10}^{2D-4}}{C_{4D-10}^{2D-4}+2 C_{3D-7}^{2D-4}-1}\nn\\
&\Leftrightarrow&
\frac{(2D-6)C_{4D-10}^{2D-4}}{(2D-6)C_{4D-10}^{2D-4}+(D-3)2 C_{3D-7}^{2D-4}}>\frac{C_{4D-10}^{2D-4}}{C_{4D-10}^{2D-4}+2 C_{3D-7}^{2D-4}-1}\nn\\ \label{mu-o-4-factor}
&\Leftrightarrow& \frac{C_{4D-10}^{2D-4}}{C_{4D-10}^{2D-4}+2 C_{3D-7}^{2D-4}-C_{3D-7}^{2D-4}}>\frac{C_{4D-10}^{2D-4}}{C_{4D-10}^{2D-4}+2 C_{3D-7}^{2D-4}-1}.
\eea
The above inequality holds since $C_{3D-7}^{2D-4}>1$. So the $(\mu^2-\o^2)$ term is positive given the bound state condition.

Secondly, let's see the $(c_D e-\o)$ term in \eqref{diff-4}, which is
\bea\label{e-o-4}
4(D-3)m^3 r_h^2(c_D e-\o)c_D e[\frac{ C_{3D-7}^{2D-4}+1}{C_{4D-10}^{2D-4}+2 C_{3D-7}^{2D-4}-1}
-\frac{ C_{3D-7}^{2D-3}}{C_{4D-10}^{2D-3}+2 C_{3D-7}^{2D-3}}]
\eea
Define the factor in the square bracket as
\bea
x_1\equiv \frac{ C_{3D-7}^{2D-4}+1}{C_{4D-10}^{2D-4}+2 C_{3D-7}^{2D-4}-1}
-\frac{ C_{3D-7}^{2D-3}}{C_{4D-10}^{2D-3}+2 C_{3D-7}^{2D-3}}.
\eea
In equation \eqref{mu-o-4-factor}, we prove that
\bea
y_1\equiv\frac{C_{4D-10}^{2D-3}}{C_{4D-10}^{2D-3}+2 C_{3D-7}^{2D-3}}-\frac{C_{4D-10}^{2D-4}}{C_{4D-10}^{2D-4}+2 C_{3D-7}^{2D-4}-1}>0.
\eea
Then we find that
\bea
y_1-2x_1=-\frac{3}{C_{4D-10}^{2D-4}+2 C_{3D-7}^{2D-4}-1}<0.
\eea
So $x_1>0$ and the $(c_D e-\o)$ term is positive given the superradiance condition.

Thirdly, let's see the $\l_l$ term in \eqref{diff-4}, which will be shown to be positive,
\bea
-4m^3\l_l[\frac{C_{5D-15}^{2D-4}+(D-6)C_{4D-12}^{2D-4}+(9-2D)C_{3D-9}^{2D-4}}{C_{4D-10}^{2D-4}+2 C_{3D-7}^{2D-4}-1}\nn\\
-\frac{C_{5D-15}^{2D-3}+(D-6)C_{4D-12}^{2D-3}+(9-2D)C_{3D-9}^{2D-3}}{C_{4D-10}^{2D-3}+2 C_{3D-7}^{2D-3}}].
\eea
The negativity of the factor in the square bracket is equivalent to
\bea
&&\frac{C_{5D-15}^{2D-4}+(D-6)C_{4D-12}^{2D-4}+(9-2D)C_{3D-9}^{2D-4}}{C_{4D-10}^{2D-4}+2 C_{3D-7}^{2D-4}-1}
<\frac{C_{5D-15}^{2D-3}+(D-6)C_{4D-12}^{2D-3}+(9-2D)C_{3D-9}^{2D-3}}{C_{4D-10}^{2D-3}+2 C_{3D-7}^{2D-3}}\nn\\ \label{l-4-factor}
&\Leftrightarrow&
\frac{C_{4D-10}^{2D-3}+2 C_{3D-7}^{2D-3}}{C_{4D-10}^{2D-4}+2 C_{3D-7}^{2D-4}-1}
<\frac{C_{5D-15}^{2D-3}+(D-6)C_{4D-12}^{2D-3}+(9-2D)C_{3D-9}^{2D-3}}{C_{5D-15}^{2D-4}+(D-6)C_{4D-12}^{2D-4}+(9-2D)C_{3D-9}^{2D-4}}.
\eea
We then consider the left and right terms of the above inequality separately. For the left term of the above inequality, we have
\bea
&&\frac{C_{4D-10}^{2D-3}+2 C_{3D-7}^{2D-3}}{C_{4D-10}^{2D-4}+2 C_{3D-7}^{2D-4}-1}=
\frac{(2D-6)C_{4D-10}^{2D-4}+(D-3)2 C_{3D-7}^{2D-4}}{C_{4D-10}^{2D-4}+2 C_{3D-7}^{2D-4}-1}\nn\\
&=&2D-6-\frac{(2D-6)(C_{3D-7}^{2D-4}-1)}{C_{4D-10}^{2D-4}+2 C_{3D-7}^{2D-4}-1}<2D-6.
\eea
For the right term in \eqref{l-4-factor}, we have
\bea
&&\frac{C_{5D-15}^{2D-3}+(D-6)C_{4D-12}^{2D-3}+(9-2D)C_{3D-9}^{2D-3}}{C_{5D-15}^{2D-4}+(D-6)C_{4D-12}^{2D-4}+(9-2D)C_{3D-9}^{2D-4}}\nn\\
&=&\frac{(3D-11)C_{5D-15}^{2D-4}+(2D-8)(D-6)C_{4D-12}^{2D-4}+(D-5)(9-2D)C_{3D-9}^{2D-3}}{C_{5D-15}^{2D-4}+(D-6)C_{4D-12}^{2D-4}+(9-2D)C_{3D-9}^{2D-4}}\nn\\
&=&2D-8+\frac{(D-3)C_{5D-15}^{2D-4}+(3-D)(9-2D)C_{3D-9}^{2D-3}}{C_{5D-15}^{2D-4}+(D-6)C_{4D-12}^{2D-4}+(9-2D)C_{3D-9}^{2D-4}}\nn\\
&>&2D-8+\frac{(D-3)C_{5D-15}^{2D-4}}{C_{5D-15}^{2D-4}+(D-6)C_{4D-12}^{2D-4}}\nn\\
&>&2D-8+\frac{(D-3)}{1+(D-6)(4/5)^8}>2D-5.
\eea
So we obtain that the right term is greater than the left term in \eqref{l-4-factor} and  the $\l_l$ term in difference \eqref{diff-4} is positive.

Finally, let's see the $D_1$ term in the difference \eqref{diff-4}, which will be shown to be positive,
\bea\label{D1-4}
-m^3D_1[\frac{C_{5D-15}^{2D-4}-5C_{4D-12}^{2D-4}+10C_{3D-9}^{2D-4}}{C_{4D-10}^{2D-4}+2 C_{3D-7}^{2D-4}-1}
-\frac{C_{5D-15}^{2D-3}-5C_{4D-12}^{2D-3}+10C_{3D-9}^{2D-3}}{C_{4D-10}^{2D-3}+2 C_{3D-7}^{2D-3}}].
\eea
The positivity of the above term is equivalent to the following inequality
\bea
&&\frac{C_{5D-15}^{2D-4}-5C_{4D-12}^{2D-4}+10C_{3D-9}^{2D-4}}{C_{4D-10}^{2D-4}+2 C_{3D-7}^{2D-4}-1}
<\frac{C_{5D-15}^{2D-3}-5C_{4D-12}^{2D-3}+10C_{3D-9}^{2D-3}}{C_{4D-10}^{2D-3}+2 C_{3D-7}^{2D-3}}\nn\\
&\Leftrightarrow&
\frac{C_{4D-10}^{2D-3}+2 C_{3D-7}^{2D-3}}{C_{4D-10}^{2D-4}+2 C_{3D-7}^{2D-4}-1}
<\frac{C_{5D-15}^{2D-3}-5C_{4D-12}^{2D-3}+10C_{3D-9}^{2D-3}}{C_{5D-15}^{2D-4}-5C_{4D-12}^{2D-4}+10C_{3D-9}^{2D-4}}\nn\\
&\Leftrightarrow&
\frac{(2D-6)C_{4D-10}^{2D-4}+(D-3)2 C_{3D-7}^{2D-4}}{C_{4D-10}^{2D-4}+2 C_{3D-7}^{2D-4}-1}\nn\\ \label{D1-4-factor}
&&<\frac{(3D-11)C_{5D-15}^{2D-4}-(2D-8)5C_{4D-12}^{2D-4}+(D-5)10C_{3D-9}^{2D-4}}{C_{5D-15}^{2D-4}-5C_{4D-12}^{2D-4}+10C_{3D-9}^{2D-4}}.
\eea
For the left term of the above inequality, we have
\bea
\frac{(2D-6)C_{4D-10}^{2D-4}+(D-3)2 C_{3D-7}^{2D-4}}{C_{4D-10}^{2D-4}+2 C_{3D-7}^{2D-4}-1}
=(2D-6)\frac{C_{4D-10}^{2D-4}+ C_{3D-7}^{2D-4}}{C_{4D-10}^{2D-4}+2 C_{3D-7}^{2D-4}-1}<2D-6.
\eea
For the right term of \eqref{D1-4-factor}, we have
\bea
&&\frac{(3D-11)C_{5D-15}^{2D-4}-(2D-8)5C_{4D-12}^{2D-4}+(D-5)10C_{3D-9}^{2D-4}}{C_{5D-15}^{2D-4}-5C_{4D-12}^{2D-4}+10C_{3D-9}^{2D-4}}\nn\\
&=&2D-8+(D-3)\frac{C_{5D-15}^{2D-4}-10C_{3D-9}^{2D-4}}{C_{5D-15}^{2D-4}-5C_{4D-12}^{2D-4}+10C_{3D-9}^{2D-4}}\nn\\
&>&2D-8+(D-3)\frac{C_{5D-15}^{2D-4}-10C_{3D-9}^{2D-4}}{C_{5D-15}^{2D-4}+10C_{3D-9}^{2D-4}}\nn\\
&=&2D-8+(D-3)\frac{1-10y}{1+10y}>2D-6.
\eea
In the last line of the above equation, we use the fact that $D>6$ and
\bea
y=\frac{C_{3D-9}^{2D-4}}{C_{5D-15}^{2D-4}}<(3/5)^{2D-4}<(3/5)^8<0.017.
\eea
So the right term of \eqref{D1-4-factor} is greater than the left term of \eqref{D1-4-factor} and the $D_1$ term in the difference \eqref{diff-4} is positive.

Then according to the positivity of the four terms in the difference \eqref{diff-4}, we obtain
\bea\label{f-7}
\text{sign}(b_{2D-4})\geqslant \text{sign}(b_{2D-3}).
\eea

\subsection{Coefficients of $z^p$, $ p=2D-4, 2D-5, 2D-6$}
In this subsection, we will consider the sign relations between the coefficients of $z^p$, $ p=2D-4, 2D-5, 2D-6$. $b_{2D-4}$ is already given in the last subsection. The left two coefficients are
\bea
b_{2D-5}&=&a_5 C_{5D-15}^{2D-5}r_h^{3D-10}+a'_4 C_{4D-10}^{2D-5}r_h^{2D-5}+a_4C_{4D-12}^{2D-5}r_h^{2D-7}
\nn\\&&+a'_3 C_{3D-7}^{2D-5}r_h^{D-2}+a_3 C_{3D-9}^{2D-5} r_h^{D-4}+a'_2 C_{2D-4}^{2D-5}r_h^{1}\nn\\
&=& r_h^{-1}(-m^3 D_1)(C_{5D-15}^{2D-5}-5C_{4D-12}^{2D-5}+10C_{3D-9}^{2D-5})\nn\\
&+&r_h^{-1}(-4m^3\l_l)(C_{5D-15}^{2D-5}+(D-6)C_{4D-12}^{2D-5}+(9-2D)C_{3D-9}^{2D-5})\nn\\
&+&4r_h(D-3)m^3(C_{4D-10}^{2D-5}+2 C_{3D-7}^{2D-5}-C_{2D-4}^{2D-5})(c_D e- \o)(\frac{ (C_{3D-7}^{2D-5}+C_{2D-4}^{2D-5}) c_D e}{C_{4D-10}^{2D-5}+2 C_{3D-7}^{2D-5}-C_{2D-4}^{2D-5}}-\o)\nn\\
&+&4r_h(D-3)m^3(\mu^2-\o^2)(2C_{3D-7}^{2D-5}-C_{4D-10}^{2D-5}-C_{2D-4}^{2D-5})
\eea
\bea
b_{2D-6}&=&a_5 C_{5D-15}^{2D-6}r_h^{3D-9}+a'_4 C_{4D-10}^{2D-6}r_h^{2D-4}+a_4C_{4D-12}^{2D-6}r_h^{2D-6}
\nn\\&&+a'_3 C_{3D-7}^{2D-6}r_h^{D-1}+a_3 C_{3D-9}^{2D-6} r_h^{D-3}+a'_2 C_{2D-4}^{2D-6}r_h^{2}
\nn\\&&+a_2C_{2D-6}^{2D-6}\nn\\
&=& (-m^3 D_1)(C_{5D-15}^{2D-6}-5C_{4D-12}^{2D-6}+10C_{3D-9}^{2D-6}-10)\nn\\
&+&(-4m^3\l_l)(C_{5D-15}^{2D-6}+(D-6)C_{4D-12}^{2D-6}+(9-2D)C_{3D-9}^{2D-6}+D-4)\nn\\
&+&4r_h^2(D-3)m^3(C_{4D-10}^{2D-6}+2 C_{3D-7}^{2D-6}-C_{2D-4}^{2D-6})(c_D e- \o)(\frac{ (C_{3D-7}^{2D-6}+C_{2D-4}^{2D-6}) c_D e}{C_{4D-10}^{2D-6}+2 C_{3D-7}^{2D-6}-C_{2D-4}^{2D-6}}-\o)\nn\\
&+&4r_h^2(D-3)m^3(\mu^2-\o^2)(2C_{3D-7}^{2D-6}-C_{4D-10}^{2D-6}-C_{2D-4}^{2D-6})
\eea

\subsubsection{$\tb_{2D-4}<\tb_{2D-5}$}
Consider the difference between two normalized coefficients, $\tb_{2D-4},\tb_{2D-5}$,
\bea\label{diff-5}
\tb_{2D-5}-\tb_{2D-4}=\frac{r_h}{C_{4D-10}^{2D-5}+2 C_{3D-7}^{2D-5}-C_{2D-4}^{2D-5}}b_{2D-5}-\tb_{2D-4}.
\eea
This difference can be decomposed into four terms and we will analyze term by term.

First, let's consider the $(\mu^2-\o^2)$ term in the difference \eqref{diff-5}, which is
\bea
4(D-3)m^3r_h^2(\mu^2-\o^2)[\frac{2C_{3D-7}^{2D-5}-C_{4D-10}^{2D-5}-C_{2D-4}^{2D-5}}{C_{4D-10}^{2D-5}+2 C_{3D-7}^{2D-5}-C_{2D-4}^{2D-5}}
-\frac{2C_{3D-7}^{2D-4}-C_{4D-10}^{2D-4}-1}{C_{4D-10}^{2D-4}+2 C_{3D-7}^{2D-4}-1}].
\eea
This term will prove to be positive. The positivity of the factor in the square bracket is equivalent to
\bea
&&\frac{2C_{3D-7}^{2D-5}-C_{4D-10}^{2D-5}-C_{2D-4}^{2D-5}}{C_{4D-10}^{2D-5}+2 C_{3D-7}^{2D-5}-C_{2D-4}^{2D-5}}
>\frac{2C_{3D-7}^{2D-4}-C_{4D-10}^{2D-4}-1}{C_{4D-10}^{2D-4}+2 C_{3D-7}^{2D-4}-1}\nn\\
&\Leftrightarrow&
\frac{-2C_{3D-7}^{2D-5}+C_{4D-10}^{2D-5}+C_{2D-4}^{2D-5}}{C_{4D-10}^{2D-5}+2 C_{3D-7}^{2D-5}-C_{2D-4}^{2D-5}}
<\frac{-2C_{3D-7}^{2D-4}+C_{4D-10}^{2D-4}+1}{C_{4D-10}^{2D-4}+2 C_{3D-7}^{2D-4}-1}\nn\\
&\Leftrightarrow&
\frac{C_{4D-10}^{2D-4}+2 C_{3D-7}^{2D-4}-1}{C_{4D-10}^{2D-5}+2 C_{3D-7}^{2D-5}-C_{2D-4}^{2D-5}}
<\frac{C_{4D-10}^{2D-4}-2C_{3D-7}^{2D-4}+1}{C_{4D-10}^{2D-5}-2C_{3D-7}^{2D-5}+C_{2D-4}^{2D-5}}\nn\\
&\Leftrightarrow&
\frac{(2D-5)C_{4D-10}^{2D-5}+(D-2)2 C_{3D-7}^{2D-5}-C_{2D-4}^{2D-5}}{C_{4D-10}^{2D-5}+2 C_{3D-7}^{2D-5}-C_{2D-4}^{2D-5}}\nn\\
&&<\frac{(2D-5)C_{4D-10}^{2D-5}-(D-2)2C_{3D-7}^{2D-5}+C_{2D-4}^{2D-5}}{C_{4D-10}^{2D-5}-2C_{3D-7}^{2D-5}+C_{2D-4}^{2D-5}}.
\eea
The left and right terms of the above inequality can be reduced as follows
\bea
&&\frac{(2D-5)C_{4D-10}^{2D-5}+(D-2)2 C_{3D-7}^{2D-5}-C_{2D-4}^{2D-5}}{C_{4D-10}^{2D-5}+2 C_{3D-7}^{2D-5}-C_{2D-4}^{2D-5}}\nn\\
&=&D-2+(D-3)\frac{C_{4D-10}^{2D-5}+C_{2D-4}^{2D-5}}{C_{4D-10}^{2D-5}+2 C_{3D-7}^{2D-5}-C_{2D-4}^{2D-5}},\\
&&\frac{(2D-5)C_{4D-10}^{2D-5}-(D-2)2C_{3D-7}^{2D-5}+C_{2D-4}^{2D-5}}{C_{4D-10}^{2D-5}-2C_{3D-7}^{2D-5}+C_{2D-4}^{2D-5}}\nn\\
&=&D-2+(D-3)\frac{C_{4D-10}^{2D-5}-C_{2D-4}^{2D-5}}{C_{4D-10}^{2D-5}-2C_{3D-7}^{2D-5}+C_{2D-4}^{2D-5}}.
\eea
Now we can get the difference between the above two terms, which is obviously positive,
\bea
(D-3)\frac{4C_{4D-10}^{2D-5}(C_{3D-7}^{2D-5}-C_{2D-4}^{2D-5})}{(C_{4D-10}^{2D-5}+2 C_{3D-7}^{2D-5}-C_{2D-4}^{2D-5})(C_{4D-10}^{2D-5}-2C_{3D-7}^{2D-5}+C_{2D-4}^{2D-5})}>0.
\eea
So the $(\mu^2-\o^2)$ term in the difference \eqref{diff-5} is positive.

Secondly, let's see the $(c_D e-\o)$ term in the difference \eqref{diff-5}, which is
\bea
4(D-3)m^3r_h^2(c_D e-\o)c_D e[\frac{C_{3D-7}^{2D-5}+C_{2D-4}^{2D-5}}{C_{4D-10}^{2D-5}+2 C_{3D-7}^{2D-5}-C_{2D-4}^{2D-5}}\nn\\
-\frac{ C_{3D-7}^{2D-4}+1}{C_{4D-10}^{2D-4}+2 C_{3D-7}^{2D-4}-1}].
\eea
We will prove its positivity in the following. Given the superradiant condition, the factor $4(D-3)m^3r_h^2(c_D e-\o)c_D e$ is positive. We just need to prove
\bea
&&\frac{C_{3D-7}^{2D-5}+C_{2D-4}^{2D-5}}{C_{4D-10}^{2D-5}+2 C_{3D-7}^{2D-5}-C_{2D-4}^{2D-5}}
>\frac{ C_{3D-7}^{2D-4}+1}{C_{4D-10}^{2D-4}+2 C_{3D-7}^{2D-4}-1}\nn\\
&\Leftrightarrow&
\frac{C_{4D-10}^{2D-4}+2 C_{3D-7}^{2D-4}-1}{C_{4D-10}^{2D-5}+2 C_{3D-7}^{2D-5}-C_{2D-4}^{2D-5}}
>\frac{ C_{3D-7}^{2D-4}+1}{C_{3D-7}^{2D-5}+C_{2D-4}^{2D-5}}\nn\\
&\Leftrightarrow&
\frac{(2D-5)C_{4D-10}^{2D-5}+(D-2)2 C_{3D-7}^{2D-5}-C_{2D-4}^{2D-5}}{C_{4D-10}^{2D-5}+2 C_{3D-7}^{2D-5}-C_{2D-4}^{2D-5}}\nn\\
&&>\frac{ (D-2)C_{3D-7}^{2D-5}+C_{2D-4}^{2D-5}}{C_{3D-7}^{2D-5}+C_{2D-4}^{2D-5}}.
\eea
It is obvious that the right term of the above inequality is less than $D-2$. For the left term of the above inequality, we have
\bea
&&\frac{(2D-5)C_{4D-10}^{2D-5}+(D-2)2 C_{3D-7}^{2D-5}-C_{2D-4}^{2D-5}}{C_{4D-10}^{2D-5}+2 C_{3D-7}^{2D-5}-C_{2D-4}^{2D-5}}\nn\\
&=&D-2+(D-3)\frac{C_{4D-10}^{2D-5}+C_{2D-4}^{2D-5}}{C_{4D-10}^{2D-5}+2 C_{3D-7}^{2D-5}-C_{2D-4}^{2D-5}}>D-2.
\eea
So the $(c_D e-\o)$ term in the difference \eqref{diff-5} is positive.

Thirdly, let's see the $\l_l$ term in the difference \eqref{diff-5}, which is
\bea
-4m^3\l_l[\frac{C_{5D-15}^{2D-5}+(D-6)C_{4D-12}^{2D-5}+(9-2D)C_{3D-9}^{2D-5}}{C_{4D-10}^{2D-5}+2 C_{3D-7}^{2D-5}-C_{2D-4}^{2D-5}}\nn\\
-\frac{C_{5D-15}^{2D-4}+(D-6)C_{4D-12}^{2D-4}+(9-2D)C_{3D-9}^{2D-4}}{C_{4D-10}^{2D-4}+2 C_{3D-7}^{2D-4}-1}].
\eea
It will be shown to be positive. The positivity of this above term is equivalent to the negativity of the factor in the square bracket, i.e.
\bea
&&\frac{C_{5D-15}^{2D-5}+(D-6)C_{4D-12}^{2D-5}+(9-2D)C_{3D-9}^{2D-5}}{C_{4D-10}^{2D-5}+2 C_{3D-7}^{2D-5}-C_{2D-4}^{2D-5}}\nn\\
&&<\frac{C_{5D-15}^{2D-4}+(D-6)C_{4D-12}^{2D-4}+(9-2D)C_{3D-9}^{2D-4}}{C_{4D-10}^{2D-4}+2 C_{3D-7}^{2D-4}-1}\nn\\
&\Leftrightarrow&
\frac{C_{4D-10}^{2D-4}+2 C_{3D-7}^{2D-4}-1}{C_{4D-10}^{2D-5}+2 C_{3D-7}^{2D-5}-C_{2D-4}^{2D-5}}
<\frac{C_{5D-15}^{2D-4}+(D-6)C_{4D-12}^{2D-4}+(9-2D)C_{3D-9}^{2D-4}}{C_{5D-15}^{2D-5}+(D-6)C_{4D-12}^{2D-5}+(9-2D)C_{3D-9}^{2D-5}}\nn\\
&\Leftrightarrow&
\frac{(2D-5)C_{4D-10}^{2D-5}+(D-2)2 C_{3D-7}^{2D-5}-C_{2D-4}^{2D-5}}{C_{4D-10}^{2D-5}+2 C_{3D-7}^{2D-5}-C_{2D-4}^{2D-5}}\nn\\ \label{l-5-factor}
&&<\frac{(3D-10)C_{5D-15}^{2D-5}+(2D-7)(D-6)C_{4D-12}^{2D-5}+(D-4)(9-2D)C_{3D-9}^{2D-5}}{C_{5D-15}^{2D-5}+(D-6)C_{4D-12}^{2D-5}+(9-2D)C_{3D-9}^{2D-5}}.
\eea
For the left term of the above inequality \eqref{l-5-factor}, we have
\bea
\frac{(2D-5)C_{4D-10}^{2D-5}+(D-2)2 C_{3D-7}^{2D-5}-C_{2D-4}^{2D-5}}{C_{4D-10}^{2D-5}+2 C_{3D-7}^{2D-5}-C_{2D-4}^{2D-5}}\nn\\
=2D-5+\frac{(6-2D) (C_{3D-7}^{2D-5}-C_{2D-4}^{2D-5})}{C_{4D-10}^{2D-5}+2 C_{3D-7}^{2D-5}-C_{2D-4}^{2D-5}}<2D-5.
\eea
For the right term of the inequality \eqref{l-5-factor}, when $D>6$, we have
\bea
&&\frac{(3D-10)C_{5D-15}^{2D-5}+(2D-7)(D-6)C_{4D-12}^{2D-5}+(D-4)(9-2D)C_{3D-9}^{2D-5}}{C_{5D-15}^{2D-5}+(D-6)C_{4D-12}^{2D-5}+(9-2D)C_{3D-9}^{2D-5}}\nn\\
&=&2D-7+\frac{(D-3)C_{5D-15}^{2D-5}+(3-D)(9-2D)C_{3D-9}^{2D-5}}{C_{5D-15}^{2D-5}+(D-6)C_{4D-12}^{2D-5}+(9-2D)C_{3D-9}^{2D-5}}\nn\\
&>&2D-7+\frac{(D-3)C_{5D-15}^{2D-5}}{C_{5D-15}^{2D-5}+(D-6)C_{4D-12}^{2D-5}}=2D-7+\frac{D-3}{1+(D-6)C_{4D-12}^{2D-5}/C_{5D-15}^{2D-5}}\nn\\
&>&2D-7+\frac{D-3}{1+(D-6)(4/5)^{2D-5}}>2D-7+3=2D-4.
\eea
So we prove that the right term is greater than the left term in the inequality \eqref{l-5-factor} and the $\l_l$ term in the difference \eqref{diff-5} is positive.

Finally, let's see the $D_1$ term in the difference \eqref{diff-5}, which is
\bea
-m^3D_1[\frac{C_{5D-15}^{2D-5}-5C_{4D-12}^{2D-5}+10C_{3D-9}^{2D-5}}{C_{4D-10}^{2D-5}+2 C_{3D-7}^{2D-5}-C_{2D-4}^{2D-5}}
-\frac{C_{5D-15}^{2D-4}-5C_{4D-12}^{2D-4}+10C_{3D-9}^{2D-4}}{C_{4D-10}^{2D-4}+2 C_{3D-7}^{2D-4}-1}].
\eea
The positivity of the above term is equivalent to the negativity of the factor in the square bracket, i.e.
\bea
&&\frac{C_{5D-15}^{2D-5}-5C_{4D-12}^{2D-5}+10C_{3D-9}^{2D-5}}{C_{4D-10}^{2D-5}+2 C_{3D-7}^{2D-5}-C_{2D-4}^{2D-5}}
<\frac{C_{5D-15}^{2D-4}-5C_{4D-12}^{2D-4}+10C_{3D-9}^{2D-4}}{C_{4D-10}^{2D-4}+2 C_{3D-7}^{2D-4}-1}\nn\\
&\Leftrightarrow&
\frac{C_{4D-10}^{2D-4}+2 C_{3D-7}^{2D-4}-1}{C_{4D-10}^{2D-5}+2 C_{3D-7}^{2D-5}-C_{2D-4}^{2D-5}}
<\frac{C_{5D-15}^{2D-4}-5C_{4D-12}^{2D-4}+10C_{3D-9}^{2D-4}}{C_{5D-15}^{2D-5}-5C_{4D-12}^{2D-5}+10C_{3D-9}^{2D-5}}\nn\\ \label{D1-5-factor}
&\Leftrightarrow&
\frac{(2D-5)C_{4D-10}^{2D-5}+(D-2)2 C_{3D-7}^{2D-5}-C_{2D-4}^{2D-5}}{C_{4D-10}^{2D-5}+2 C_{3D-7}^{2D-5}-C_{2D-4}^{2D-5}}\nn\\ \label{D1-5-factor}
&&<\frac{(3D-10)C_{5D-15}^{2D-5}-(2D-7)5C_{4D-12}^{2D-5}+(D-4)10C_{3D-9}^{2D-5}}{C_{5D-15}^{2D-5}-5C_{4D-12}^{2D-5}+10C_{3D-9}^{2D-5}}
\eea
For the left term of the above inequality \eqref{D1-5-factor}, we have
\bea
\frac{(2D-5)C_{4D-10}^{2D-5}+(D-2)2 C_{3D-7}^{2D-5}-C_{2D-4}^{2D-5}}{C_{4D-10}^{2D-5}+2 C_{3D-7}^{2D-5}-C_{2D-4}^{2D-5}}\nn\\
=2D-5+\frac{(6-2D)(C_{3D-7}^{2D-5}-C_{2D-4}^{2D-5})}{C_{4D-10}^{2D-5}+2 C_{3D-7}^{2D-5}-C_{2D-4}^{2D-5}}<2D-5.
\eea
For the right term of the inequality \eqref{D1-5-factor}, we have
\bea
&&\frac{(3D-10)C_{5D-15}^{2D-5}-(2D-7)5C_{4D-12}^{2D-5}+(D-4)10C_{3D-9}^{2D-5}}{C_{5D-15}^{2D-5}-5C_{4D-12}^{2D-5}+10C_{3D-9}^{2D-5}}\nn\\
&=&2D-7+\frac{(D-3)(C_{5D-15}^{2D-5}-10C_{3D-9}^{2D-5})}{C_{5D-15}^{2D-5}-5C_{4D-12}^{2D-5}+10C_{3D-9}^{2D-5}}\nn\\
&>&2D-7+\frac{(D-3)(C_{5D-15}^{2D-5}-10C_{3D-9}^{2D-5})}{C_{5D-15}^{2D-5}+10C_{3D-9}^{2D-5}}\nn\\
&>&2D-7+(D-3)\frac{1-10C_{3D-9}^{2D-5}/C_{5D-15}^{2D-5}}{1+10C_{3D-9}^{2D-5}/C_{5D-15}^{2D-5}}\nn\\
&>&2D-7+(D-3)\frac{1-10(3/5)^{2D-5}}{1+10(3/5)^{2D-5}}>2D-7+3=2D-4.
\eea
So we prove that the right term is greater than the left term in the inequality \eqref{D1-5-factor}.

According to the positivity of the four terms in the difference \eqref{diff-5}, we obtain
\bea\label{f-8}
\text{sign}(b_{2D-5})\geqslant \text{sign}(b_{2D-4}).
\eea

\subsubsection{$\tb_{2D-5}<\tb_{2D-6}$}

Now, consider the difference between two normalized coefficients, $\tb_{2D-6},\tb_{2D-5}$,
\bea\label{diff-6}
\tb_{2D-6}-\tb_{2D-5}=\frac{b_{2D-6}}{C_{4D-10}^{2D-6}+2 C_{3D-7}^{2D-6}-C_{2D-4}^{2D-6}}-\tb_{2D-5}
\eea
The difference can be decomposed into four terms and we will analyze term by term.

First, let's see the $(\mu^2-\o^2)$ term in the difference \eqref{diff-6},which is
\bea
4(D-3)m^3 r_h^2(\mu^2-\o^2)[\frac{2C_{3D-7}^{2D-6}-C_{4D-10}^{2D-6}-C_{2D-4}^{2D-6}}{C_{4D-10}^{2D-6}+2 C_{3D-7}^{2D-6}-C_{2D-4}^{2D-6}}\nn\\
-\frac{2C_{3D-7}^{2D-5}-C_{4D-10}^{2D-5}-C_{2D-4}^{2D-5}}{C_{4D-10}^{2D-5}+2 C_{3D-7}^{2D-5}-C_{2D-4}^{2D-5}}]
\eea
The positivity of this term is equivalent to the positivity of the factor in the square bracket, i.e.
\bea
&&\frac{2C_{3D-7}^{2D-6}-C_{4D-10}^{2D-6}-C_{2D-4}^{2D-6}}{C_{4D-10}^{2D-6}+2 C_{3D-7}^{2D-6}-C_{2D-4}^{2D-6}}
>\frac{2C_{3D-7}^{2D-5}-C_{4D-10}^{2D-5}-C_{2D-4}^{2D-5}}{C_{4D-10}^{2D-5}+2 C_{3D-7}^{2D-5}-C_{2D-4}^{2D-5}}\nn\\
&\Leftrightarrow&
\frac{-2C_{3D-7}^{2D-6}+C_{4D-10}^{2D-6}+C_{2D-4}^{2D-6}}{C_{4D-10}^{2D-6}+2 C_{3D-7}^{2D-6}-C_{2D-4}^{2D-6}}
<\frac{-2C_{3D-7}^{2D-5}+C_{4D-10}^{2D-5}+C_{2D-4}^{2D-5}}{C_{4D-10}^{2D-5}+2 C_{3D-7}^{2D-5}-C_{2D-4}^{2D-5}}\nn\\
&\Leftrightarrow&
\frac{C_{4D-10}^{2D-5}+2 C_{3D-7}^{2D-5}-C_{2D-4}^{2D-5}}{C_{4D-10}^{2D-6}+2 C_{3D-7}^{2D-6}-C_{2D-4}^{2D-6}}
<\frac{C_{4D-10}^{2D-5}-2C_{3D-7}^{2D-5}+C_{2D-4}^{2D-5}}{C_{4D-10}^{2D-6}-2C_{3D-7}^{2D-6}+C_{2D-4}^{2D-6}}\nn\\
&\Leftrightarrow&
\frac{(2D-4)C_{4D-10}^{2D-6}+(D-1)2 C_{3D-7}^{2D-4}-2C_{2D-4}^{2D-6}}{C_{4D-10}^{2D-6}+2 C_{3D-7}^{2D-6}-C_{2D-4}^{2D-6}}\nn\\ \label{mu-o-6-factor}
&&<\frac{(2D-4)C_{4D-10}^{2D-6}-(D-1)2C_{3D-7}^{2D-6}+2C_{2D-4}^{2D-6}}{C_{4D-10}^{2D-6}-2C_{3D-7}^{2D-6}+C_{2D-4}^{2D-6}}.
\eea
For the left term of the above inequality, we have
\bea
\frac{(2D-4)C_{4D-10}^{2D-6}+(D-1)2 C_{3D-7}^{2D-4}-2C_{2D-4}^{2D-6}}{C_{4D-10}^{2D-6}+2 C_{3D-7}^{2D-6}-C_{2D-4}^{2D-6}}\nn\\
=D-1+\frac{(D-3)C_{4D-10}^{2D-6}+(D-3)C_{2D-4}^{2D-6}}{C_{4D-10}^{2D-6}+2 C_{3D-7}^{2D-6}-C_{2D-4}^{2D-6}}.
\eea
For the right term of the inequality\eqref{mu-o-6-factor}, we have
\bea
\frac{(2D-4)C_{4D-10}^{2D-6}-(D-1)2C_{3D-7}^{2D-6}+2C_{2D-4}^{2D-6}}{C_{4D-10}^{2D-6}-2C_{3D-7}^{2D-6}+C_{2D-4}^{2D-6}}\nn\\
=D-1+\frac{(D-3)C_{4D-10}^{2D-6}-(D-3)C_{2D-4}^{2D-6}}{C_{4D-10}^{2D-6}-2C_{3D-7}^{2D-6}+C_{2D-4}^{2D-6}}.
\eea
Then the difference between the right term and the left term of the inequality \eqref{mu-o-6-factor} is
\bea
\frac{4(D-3)C_{4D-10}^{2D-6}(C_{3D-7}^{2D-6}-C_{2D-4}^{2D-6})}{(C_{4D-10}^{2D-6}+2 C_{3D-7}^{2D-6}-C_{2D-4}^{2D-6})(C_{4D-10}^{2D-6}-2C_{3D-7}^{2D-6}+C_{2D-4}^{2D-6})},
\eea
which is obviously positive. So the $(\mu^2-\o^2)$ term in the difference \eqref{diff-6} is positive.

Secondly, let's see the $(c_D e-\o)$ term in the difference \eqref{diff-6},which is
\bea
4(D-3)m^3r_h^2(c_D e-\o)c_D e[\frac{C_{3D-7}^{2D-6}+C_{2D-4}^{2D-6}}{C_{4D-10}^{2D-6}+2 C_{3D-7}^{2D-6}-C_{2D-4}^{2D-6}}\nn\\
-\frac{C_{3D-7}^{2D-5}+C_{2D-4}^{2D-5}}{C_{4D-10}^{2D-5}+2 C_{3D-7}^{2D-5}-C_{2D-4}^{2D-5}}].
\eea
The positivity of the above term is equivalent to the positivity of the factor in the square bracket, i.e.
\bea
&&\frac{C_{3D-7}^{2D-6}+C_{2D-4}^{2D-6}}{C_{4D-10}^{2D-6}+2 C_{3D-7}^{2D-6}-C_{2D-4}^{2D-6}}
>\frac{C_{3D-7}^{2D-5}+C_{2D-4}^{2D-5}}{C_{4D-10}^{2D-5}+2 C_{3D-7}^{2D-5}-C_{2D-4}^{2D-5}}\nn\\
&\Leftrightarrow&
\frac{C_{4D-10}^{2D-5}+2 C_{3D-7}^{2D-5}-C_{2D-4}^{2D-5}}{C_{4D-10}^{2D-6}+2 C_{3D-7}^{2D-6}-C_{2D-4}^{2D-6}}
>\frac{C_{3D-7}^{2D-5}+C_{2D-4}^{2D-5}}{C_{3D-7}^{2D-6}+C_{2D-4}^{2D-6}}\nn\\
&\Leftrightarrow&
\frac{(2D-4)C_{4D-10}^{2D-6}+(D-1)2 C_{3D-7}^{2D-6}-2C_{2D-4}^{2D-6}}{C_{4D-10}^{2D-6}+2 C_{3D-7}^{2D-6}-C_{2D-4}^{2D-6}}\nn\\ \label{e-o-6-factor}
&&>\frac{(D-1)C_{3D-7}^{2D-6}+2C_{2D-4}^{2D-6}}{C_{3D-7}^{2D-6}+C_{2D-4}^{2D-6}}.
\eea
For the right term of the above inequality, when $D>6$, it is easy to see that
\bea
\frac{(D-1)C_{3D-7}^{2D-6}+2C_{2D-4}^{2D-6}}{C_{3D-7}^{2D-6}+C_{2D-4}^{2D-6}}<D-1.
\eea
For the left term of the inequality \eqref{e-o-6-factor}, we have
\bea
\frac{(2D-4)C_{4D-10}^{2D-6}+(D-1)2 C_{3D-7}^{2D-6}-2C_{2D-4}^{2D-6}}{C_{4D-10}^{2D-6}+2 C_{3D-7}^{2D-6}-C_{2D-4}^{2D-6}}\nn\\
=D-1+\frac{(D-3)C_{4D-10}^{2D-6}+(D-3)C_{2D-4}^{2D-6}}{C_{4D-10}^{2D-6}+2 C_{3D-7}^{2D-6}-C_{2D-4}^{2D-6}}>D-1.
\eea
So the left term is greater than the right term of the inequality \eqref{e-o-6-factor}. The $(c_D e-\o)$ term in the difference \eqref{diff-6} is positive.

Thirdly, let's see the $\l_l$ term in the difference \eqref{diff-6},which is
\bea
-4m^3\l_l[\frac{C_{5D-15}^{2D-6}+(D-6)C_{4D-12}^{2D-6}+(9-2D)C_{3D-9}^{2D-6}+D-4}{C_{4D-10}^{2D-6}+2 C_{3D-7}^{2D-6}-C_{2D-4}^{2D-6}}\nn\\
-\frac{C_{5D-15}^{2D-5}+(D-6)C_{4D-12}^{2D-5}+(9-2D)C_{3D-9}^{2D-5}}{C_{4D-10}^{2D-5}+2 C_{3D-7}^{2D-5}-C_{2D-4}^{2D-5}}].
\eea
The positivity of the above term is equivalent to the negativity of the factor in the square bracket, i.e.
\bea
&&\frac{C_{5D-15}^{2D-6}+(D-6)C_{4D-12}^{2D-6}+(9-2D)C_{3D-9}^{2D-6}+D-4}{C_{4D-10}^{2D-6}+2 C_{3D-7}^{2D-6}-C_{2D-4}^{2D-6}}\nn\\
&&<\frac{C_{5D-15}^{2D-5}+(D-6)C_{4D-12}^{2D-5}+(9-2D)C_{3D-9}^{2D-5}}{C_{4D-10}^{2D-5}+2 C_{3D-7}^{2D-5}-C_{2D-4}^{2D-5}}\nn\\
&\Leftrightarrow&
\frac{C_{4D-10}^{2D-5}+2 C_{3D-7}^{2D-5}-C_{2D-4}^{2D-5}}{C_{4D-10}^{2D-6}+2 C_{3D-7}^{2D-6}-C_{2D-4}^{2D-6}}
<\frac{C_{5D-15}^{2D-5}+(D-6)C_{4D-12}^{2D-5}+(9-2D)C_{3D-9}^{2D-5}}{C_{5D-15}^{2D-6}+(D-6)C_{4D-12}^{2D-6}+(9-2D)C_{3D-9}^{2D-6}+D-4}\nn\\
&\Leftrightarrow&
\frac{(2D-4)C_{4D-10}^{2D-6}+(D-1)2 C_{3D-7}^{2D-6}-2C_{2D-4}^{2D-6}}{C_{4D-10}^{2D-6}+2 C_{3D-7}^{2D-6}-C_{2D-4}^{2D-6}}\nn\\ \label{l-6-factor}
&&<\frac{(3D-9)C_{5D-15}^{2D-6}+(2D-6)(D-6)C_{4D-12}^{2D-6}+(D-3)(9-2D)C_{3D-9}^{2D-6}}{C_{5D-15}^{2D-6}+(D-6)C_{4D-12}^{2D-6}+(9-2D)C_{3D-9}^{2D-6}+D-4}
\eea
For the left term of the above inequality, we have
\bea
\frac{(2D-4)C_{4D-10}^{2D-6}+(D-1)2 C_{3D-7}^{2D-6}-2C_{2D-4}^{2D-6}}{C_{4D-10}^{2D-6}+2 C_{3D-7}^{2D-6}-C_{2D-4}^{2D-6}}\nn\\
=2D-4+\frac{(6-2D)(C_{3D-7}^{2D-6}-C_{2D-4}^{2D-6})}{C_{4D-10}^{2D-6}+2 C_{3D-7}^{2D-6}-C_{2D-4}^{2D-6}}<2D-4.
\eea
For the right term of the inequality \eqref{l-6-factor}, we have
\bea
&&\frac{(3D-9)C_{5D-15}^{2D-6}+(2D-6)(D-6)C_{4D-12}^{2D-6}+(D-3)(9-2D)C_{3D-9}^{2D-6}}{C_{5D-15}^{2D-6}+(D-6)C_{4D-12}^{2D-6}+(9-2D)C_{3D-9}^{2D-6}+D-4}\nn\\
&=&2D-6+\frac{(D-3)C_{5D-15}^{2D-6}+(D-3)(2D-9)C_{3D-9}^{2D-6}-(2D-6)(D-4)}{C_{5D-15}^{2D-6}+(D-6)C_{4D-12}^{2D-6}+(9-2D)C_{3D-9}^{2D-6}+D-4}\nn\\
&>&2D-6+\frac{(D-3)C_{5D-15}^{2D-6}}{C_{5D-15}^{2D-6}+(D-6)C_{4D-12}^{2D-6}}>2D-6+\frac{D-3}{1+(D-6)(4/5)^{2D-6}}\nn\\
&>&2D-6+3=2D-3.
\eea
So the right term is greater than the left term of the inequality \eqref{l-6-factor}. The $\l_l$ term in the difference \eqref{diff-6} is positive.

Finally, let see the $D_1$ term in the difference \eqref{diff-6},which is
\bea
-m^3D_1[\frac{C_{5D-15}^{2D-6}-5C_{4D-12}^{2D-6}+10C_{3D-9}^{2D-6}-10}{C_{4D-10}^{2D-6}+2 C_{3D-7}^{2D-6}-C_{2D-4}^{2D-6}}\nn\\
-\frac{C_{5D-15}^{2D-5}-5C_{4D-12}^{2D-5}+10C_{3D-9}^{2D-5}}{C_{4D-10}^{2D-5}+2 C_{3D-7}^{2D-5}-C_{2D-4}^{2D-5}}].
\eea
The positivity of the above term is equivalent to the negativity of the factor in the square bracket, i.e.
\bea
&&\frac{C_{5D-15}^{2D-6}-5C_{4D-12}^{2D-6}+10C_{3D-9}^{2D-6}-10}{C_{4D-10}^{2D-6}+2 C_{3D-7}^{2D-6}-C_{2D-4}^{2D-6}}
<\frac{C_{5D-15}^{2D-5}-5C_{4D-12}^{2D-5}+10C_{3D-9}^{2D-5}}{C_{4D-10}^{2D-5}+2 C_{3D-7}^{2D-5}-C_{2D-4}^{2D-5}}\nn\\
&\Leftrightarrow&
\frac{C_{4D-10}^{2D-5}+2 C_{3D-7}^{2D-5}-C_{2D-4}^{2D-5}}{C_{4D-10}^{2D-6}+2 C_{3D-7}^{2D-6}-C_{2D-4}^{2D-6}}
<\frac{C_{5D-15}^{2D-5}-5C_{4D-12}^{2D-5}+10C_{3D-9}^{2D-5}}{C_{5D-15}^{2D-6}-5C_{4D-12}^{2D-6}+10C_{3D-9}^{2D-6}-10}\nn\\
&\Leftrightarrow&
\frac{(2D-4)C_{4D-10}^{2D-6}+(D-1)2 C_{3D-7}^{2D-6}-2C_{2D-4}^{2D-6}}{C_{4D-10}^{2D-6}+2 C_{3D-7}^{2D-6}-C_{2D-4}^{2D-6}}\nn\\
&&<\frac{(3D-9)C_{5D-15}^{2D-6}-(2D-6)5C_{4D-12}^{2D-6}+(D-3)10C_{3D-9}^{2D-6}}{C_{5D-15}^{2D-6}-5C_{4D-12}^{2D-6}+10C_{3D-9}^{2D-6}-10}
.\label{D1-6-factor}
\eea
For the left term of the above inequality, we have
\bea
&&\frac{(2D-4)C_{4D-10}^{2D-6}+(D-1)2 C_{3D-7}^{2D-6}-2C_{2D-4}^{2D-6}}{C_{4D-10}^{2D-6}+2 C_{3D-7}^{2D-6}-C_{2D-4}^{2D-6}}\nn\\
&=&2D-4+\frac{(6-2D) (C_{3D-7}^{2D-6}-C_{2D-4}^{2D-6})}{C_{4D-10}^{2D-6}+2 C_{3D-7}^{2D-6}-C_{2D-4}^{2D-6}}<2D-4.
\eea
For the right term of the inequality \eqref{D1-6-factor}, we have
\bea
&&\frac{(3D-9)C_{5D-15}^{2D-6}-(2D-6)5C_{4D-12}^{2D-6}+(D-3)10C_{3D-9}^{2D-6}}{C_{5D-15}^{2D-6}-5C_{4D-12}^{2D-6}+10C_{3D-9}^{2D-6}-10}\nn\\
&=&2D-6+\frac{(D-3)C_{5D-15}^{2D-6}+(3-D)10C_{3D-9}^{2D-6}+10(2D-6)}{C_{5D-15}^{2D-6}-5C_{4D-12}^{2D-6}+10C_{3D-9}^{2D-6}-10}\nn\\
&>&2D-6+(D-3)\frac{C_{5D-15}^{2D-6}-10C_{3D-9}^{2D-6}}{C_{5D-15}^{2D-6}+10C_{3D-9}^{2D-6}}\nn\\
&>&2D-6+(D-3)\frac{1-10(3/5)^{2D-6}}{1+10(3/5)^{2D-6}}>2D-4.
\eea
So the right term is greater than the left term of the inequality \eqref{D1-6-factor}. The $D_1$ term in the difference \eqref{diff-6} is positive.

Based on the positivity of the four terms in the difference \eqref{diff-6}, we obtain
\bea\label{f-9}
\text{sign}(b_{2D-6})\geqslant \text{sign}(b_{2D-5}).
\eea

\subsection{Coefficients of $z^p$, $ 2D-6> p>D-3$}
Since $D>6(\geqslant 7)$, we have $p\geqslant D-2\geqslant 5$ in this case. For $ C_{5D-15}^{p}, C_{4D-12}^{p}, C_{3D-9}^{p} $, we have the following inequalities
\bea
&&\frac{C_{3D-9}^{p}}{C_{5D-15}^{p}}<(3/5)^p<(3/5)^5\Rightarrow C_{5D-15}^{p}>(5/3)^5C_{3D-9}^{p}>12 C_{3D-9}^{p},\nn\\ \label{c5c4c3}
&&\frac{C_{3D-9}^{p}}{C_{4D-12}^{p}}<(3/4)^p<(3/4)^5\Rightarrow C_{4D-12}^{p}>(4/3)^5C_{3D-9}^{p}>4 C_{3D-9}^{p}.
\eea
The coefficient of $z^p$ when $ 2D-6> p>D-3$ is
\bea
b_{p}&=&a_5 C_{5D-15}^{p}r_h^{5D-15-p}+a'_4 C_{4D-10}^{p}r_h^{4D-10-p}+a_4C_{4D-12}^{p}r_h^{4D-12-p}
\nn\\&&+a'_3 C_{3D-7}^{p}r_h^{3D-7-p}+a_3 C_{3D-9}^{p} r_h^{3D-9-p}+a'_2 C_{2D-4}^{p}r_h^{2D-4-p}
\nn\\&&+a_2C_{2D-6}^{p}r_h^{2D-6-p}\nn\\
&=& r_h^{-p}(-m^5 D_1)(C_{5D-15}^{p}-5C_{4D-12}^{p}+10C_{3D-9}^p-10C_{2D-6}^{p})\nn\\
&+&r_h^{-p}(-4m^5\l_l)(C_{5D-15}^{p}+(D-6)C_{4D-12}^{p}+(9-2D)C_{3D-9}^p+(D-4)C_{2D-6}^{p})\nn\\
&+&r_h^{-p+2}4(D-3)m^5(C_{4D-10}^{p}+2 C_{3D-7}^{p}-C_{2D-4}^{p})(c_D e- \o)(\frac{(C_{3D-7}^{p}+C_{2D-4}^{p}) c_D e}{C_{4D-10}^{p}+2 C_{3D-7}^{p}-C_{2D-4}^{p}}-\o)\nn\\
&+&r_h^{-p+2}4(D-3)m^5(\mu^2-\o^2)(2C_{3D-7}^{p}-C_{4D-10}^{p}-C_{2D-4}^{p}).
\eea

Now we consider the difference between two normalized coefficients, $\tb_{p},\tb_{p+1}$,
\bea\label{diff-7}
\tb_{p}-\tb_{p+1}=\frac{r_h^p}{C_{4D-10}^{p}+2 C_{3D-7}^{p}-C_{2D-4}^{p}}b_{p}
-\frac{r_h^{p+1}}{C_{4D-10}^{p+1}+2 C_{3D-7}^{p+1}-C_{2D-4}^{p+1}}b_{p+1}
\eea
The difference can be decomposed into four terms and we will analyze term by term.

First, let's see the $(\mu^2-\o^2)$ term in the difference \eqref{diff-7}, which is
\bea
4(D-3)m^5r_h^2(\mu^2-\o^2)[\frac{2C_{3D-7}^{p}-C_{4D-10}^{p}-C_{2D-4}^{p}}{C_{4D-10}^{p}+2 C_{3D-7}^{p}-C_{2D-4}^{p}}\nn\\
-\frac{2C_{3D-7}^{p+1}-C_{4D-10}^{p+1}-C_{2D-4}^{p+1}}{C_{4D-10}^{p+1}+2 C_{3D-7}^{p+1}-C_{2D-4}^{p+1}}].
\eea
The positivity of the above term is equivalent to the positivity of the factor in the square bracket, i.e.
\bea
&&\frac{2C_{3D-7}^{p}-C_{4D-10}^{p}-C_{2D-4}^{p}}{C_{4D-10}^{p}+2 C_{3D-7}^{p}-C_{2D-4}^{p}}
>\frac{2C_{3D-7}^{p+1}-C_{4D-10}^{p+1}-C_{2D-4}^{p+1}}{C_{4D-10}^{p+1}+2 C_{3D-7}^{p+1}-C_{2D-4}^{p+1}}\nn\\
&\Leftrightarrow&
\frac{C_{4D-10}^{p+1}+2 C_{3D-7}^{p+1}-C_{2D-4}^{p+1}}{C_{4D-10}^{p}+2 C_{3D-7}^{p}-C_{2D-4}^{p}}
<\frac{C_{4D-10}^{p+1}-2C_{3D-7}^{p+1}+C_{2D-4}^{p+1}}{C_{4D-10}^{p}-2C_{3D-7}^{p}+C_{2D-4}^{p}}\nn\\
&\Leftrightarrow&
\frac{(4D-p-10)C_{4D-10}^{p}+(3D-p-7)2 C_{3D-7}^{p}-(2D-p-4)C_{2D-4}^{p}}{C_{4D-10}^{p}+2 C_{3D-7}^{p}-C_{2D-4}^{p}}\nn\\
&&<\frac{(4D-p-10)C_{4D-10}^{p}-(3D-p-7)2C_{3D-7}^{p}+(2D-p-4)C_{2D-4}^{p}}{C_{4D-10}^{p}-2C_{3D-7}^{p}+C_{2D-4}^{p}}
\eea
After a straightforward calculation, the difference of the right and left terms of the above inequality is
\bea\nn
\frac{4(D-3)C_{4D-10}^{p}( C_{3D-7}^{p}-C_{2D-4}^{p})}{(C_{4D-10}^{p}+2 C_{3D-7}^{p}-C_{2D-4}^{p})(C_{4D-10}^{p}-2C_{3D-7}^{p}+C_{2D-4}^{p})}>0.
\eea
So the $(\mu^2-\o^2)$ term in the difference \eqref{diff-7} is positive.

Secondly, let's see the $(c_D e-\o)$ term in the difference \eqref{diff-7}, which is
\bea
4(D-3)m^5r_h^2(c_D e-\o)c_D e[\frac{C_{3D-7}^{p}+C_{2D-4}^{p}}{C_{4D-10}^{p}+2 C_{3D-7}^{p}-C_{2D-4}^{p}}\nn\\
-\frac{C_{3D-7}^{p+1}+C_{2D-4}^{p+1}}{C_{4D-10}^{p+1}+2 C_{3D-7}^{p+1}-C_{2D-4}^{p+1}}]
\eea
The positivity of the above term is equivalent to the positivity of the factor in the square bracket, i.e.
\bea
&&\frac{C_{3D-7}^{p}+C_{2D-4}^{p}}{C_{4D-10}^{p}+2 C_{3D-7}^{p}-C_{2D-4}^{p}}>
\frac{C_{3D-7}^{p+1}+C_{2D-4}^{p+1}}{C_{4D-10}^{p+1}+2 C_{3D-7}^{p+1}-C_{2D-4}^{p+1}}\nn\\
&\Leftrightarrow&
\frac{(4D-p-10)C_{4D-10}^{p}+(3D-p-7)2 C_{3D-7}^{p}-(2D-p-4)C_{2D-4}^{p}}{C_{4D-10}^{p}+2 C_{3D-7}^{p}-C_{2D-4}^{p}}\nn\\ \label{e-o-7-factor}
&&>\frac{(3D-p-7)C_{3D-7}^{p}+(2D-p-4)C_{2D-4}^{p}}{C_{3D-7}^{p}+C_{2D-4}^{p}}.
\eea
For the right term of the above inequality, it is easy to see that
\bea
\frac{(3D-p-7)C_{3D-7}^{p}+(2D-p-4)C_{2D-4}^{p}}{C_{3D-7}^{p}+C_{2D-4}^{p}}<3D-p-7.
\eea
For the left term of the inequality \eqref{e-o-7-factor}, we have
\bea
&&\frac{(4D-p-10)C_{4D-10}^{p}+(3D-p-7)2 C_{3D-7}^{p}-(2D-p-4)C_{2D-4}^{p}}{C_{4D-10}^{p}+2 C_{3D-7}^{p}-C_{2D-4}^{p}}\nn\\
&=&3D-p-7+\frac{(D-3)(C_{4D-10}^{p}+C_{2D-4}^{p})}{C_{4D-10}^{p}+2 C_{3D-7}^{p}-C_{2D-4}^{p}}>3D-p-7.
\eea
So the left term is greater than the right term of the inequality \eqref{e-o-7-factor}. The $(c_D e-\o)$ term in the difference \eqref{diff-7} is positive.

Thirdly, let's see the $\l_l$ term in the difference \eqref{diff-7}, which is
\bea
-4m^5\l_l[\frac{C_{5D-15}^{p}+(D-6)C_{4D-12}^{p}+(9-2D)C_{3D-9}^p+(D-4)C_{2D-6}^{p}}{C_{4D-10}^{p}+2 C_{3D-7}^{p}-C_{2D-4}^{p}}\nn\\
-\frac{C_{5D-15}^{p+1}+(D-6)C_{4D-12}^{p+1}+(9-2D)C_{3D-9}^{p+1}+(D-4)C_{2D-6}^{p+1}}{C_{4D-10}^{p+1}+2 C_{3D-7}^{p+1}-C_{2D-4}^{p+1}}].
\eea
The positivity of the above term is equivalent to the negativity of the factor in the square bracket, i.e.
\bea
&&\frac{C_{5D-15}^{p}+(D-6)C_{4D-12}^{p}+(9-2D)C_{3D-9}^p+(D-4)C_{2D-6}^{p}}{C_{4D-10}^{p}+2 C_{3D-7}^{p}-C_{2D-4}^{p}}\nn\\
&&<\frac{C_{5D-15}^{p+1}+(D-6)C_{4D-12}^{p+1}+(9-2D)C_{3D-9}^{p+1}+(D-4)C_{2D-6}^{p+1}}{C_{4D-10}^{p+1}+2 C_{3D-7}^{p+1}-C_{2D-4}^{p+1}}\nn\\
&\Leftrightarrow&
\frac{C_{4D-10}^{p+1}+2 C_{3D-7}^{p+1}-C_{2D-4}^{p+1}}{C_{4D-10}^{p}+2 C_{3D-7}^{p}-C_{2D-4}^{p}}\nn\\
&&<\frac{C_{5D-15}^{p+1}+(D-6)C_{4D-12}^{p+1}+(9-2D)C_{3D-9}^{p+1}+(D-4)C_{2D-6}^{p+1}}{C_{5D-15}^{p}+(D-6)C_{4D-12}^{p}+(9-2D)C_{3D-9}^p+(D-4)C_{2D-6}^{p}}\nn\\
&\Leftrightarrow&
\frac{(4D-p-10)C_{4D-10}^{p}+(3D-p-7)2 C_{3D-7}^{p}-(2D-p-4)C_{2D-4}^{p}}{C_{4D-10}^{p}+2 C_{3D-7}^{p}-C_{2D-4}^{p}}\nn\\ \label{l-7-factor}
&&<\frac{a_5C_{5D-15}^{p}+a_4(D-6)C_{4D-12}^{p}+a_3(9-2D)C_{3D-9}^{p}+a_2(D-4)C_{2D-6}^{p}}{C_{5D-15}^{p}+(D-6)C_{4D-12}^{p}+(9-2D)C_{3D-9}^p+(D-4)C_{2D-6}^{p}}.
\eea
In the above inequality, $a_i=(D-3)i-p$. For the left term of the above inequality, we have
\bea
&&\frac{(4D-p-10)C_{4D-10}^{p}+(3D-p-7)2 C_{3D-7}^{p}-(2D-p-4)C_{2D-4}^{p}}{C_{4D-10}^{p}+2 C_{3D-7}^{p}-C_{2D-4}^{p}}\nn\\
&=&4D-p-10+\frac{(6-2D)( C_{3D-7}^{p}-C_{2D-4}^{p})}{C_{4D-10}^{p}+2 C_{3D-7}^{p}-C_{2D-4}^{p}}<4D-p-10.
\eea
For the right term of the inequality \eqref{l-7-factor}, we have
\bea
&&\frac{a_5C_{5D-15}^{p}+a_4(D-6)C_{4D-12}^{p}+a_3(9-2D)C_{3D-9}^{p}+a_2(D-4)C_{2D-6}^{p}}{C_{5D-15}^{p}+(D-6)C_{4D-12}^{p}+(9-2D)C_{3D-9}^p+(D-4)C_{2D-6}^{p}}\nn\\
&=&a_4+\frac{(D-3)C_{5D-15}^{p}+(3-D)(9-2D)C_{3D-9}^{p}+2(3-D)(D-4)C_{2D-6}^{p}}{C_{5D-15}^{p}+(D-6)C_{4D-12}^{p}+(9-2D)C_{3D-9}^p+(D-4)C_{2D-6}^{p}}\nn\\
&>&4D-p-12+\frac{(D-3)C_{5D-15}^{p}}{C_{5D-15}^{p}+(D-6)C_{4D-12}^{p}}\nn\\
&>&4D-p-12+\frac{D-3}{1+(D-6)(4/5)^{4}}>4D-p-10.
\eea
So the right term is greater than the left term of the inequality \eqref{l-7-factor}. The $\l_l$ term in the difference \eqref{diff-7} is positive.

Finally, let's see the $D_1$ term in the difference \eqref{diff-7}, which is
\bea
-m^5D_1[\frac{C_{5D-15}^{p}-5C_{4D-12}^{p}+10C_{3D-9}^p-10C_{2D-6}^{p}}{C_{4D-10}^{p}+2 C_{3D-7}^{p}-C_{2D-4}^{p}}\nn\\
-\frac{C_{5D-15}^{p+1}-5C_{4D-12}^{p+1}+10C_{3D-9}^{p+1}-10C_{2D-6}^{p+1}}{C_{4D-10}^{p+1}+2 C_{3D-7}^{p+1}-C_{2D-4}^{p+1}}].
\eea
The positivity of the above term is equivalent to the negativity of the factor in the square bracket, i.e.
\bea
&&\frac{C_{5D-15}^{p}-5C_{4D-12}^{p}+10C_{3D-9}^p-10C_{2D-6}^{p}}{C_{4D-10}^{p}+2 C_{3D-7}^{p}-C_{2D-4}^{p}}
<\frac{C_{5D-15}^{p+1}-5C_{4D-12}^{p+1}+10C_{3D-9}^{p+1}-10C_{2D-6}^{p+1}}{C_{4D-10}^{p+1}+2 C_{3D-7}^{p+1}-C_{2D-4}^{p+1}}\nn\\
&&\frac{C_{4D-10}^{p+1}+2 C_{3D-7}^{p+1}-C_{2D-4}^{p+1}}{C_{4D-10}^{p}+2 C_{3D-7}^{p}-C_{2D-4}^{p}}
<\frac{C_{5D-15}^{p+1}-5C_{4D-12}^{p+1}+10C_{3D-9}^{p+1}-10C_{2D-6}^{p+1}}{C_{5D-15}^{p}-5C_{4D-12}^{p}+10C_{3D-9}^p-10C_{2D-6}^{p}}\nn\\
&\Leftrightarrow&
\frac{(4D-p-10)C_{4D-10}^{p}+(3D-p-7)2 C_{3D-7}^{p}-(2D-p-4)C_{2D-4}^{p}}{C_{4D-10}^{p}+2 C_{3D-7}^{p}-C_{2D-4}^{p}}\nn\\ \label{D1-7-factor}
&&<\frac{a_5C_{5D-15}^{p}-a_45C_{4D-12}^{p}+a_310C_{3D-9}^{p}-a_210C_{2D-6}^{p}}{C_{5D-15}^{p}-5C_{4D-12}^{p}+10C_{3D-9}^p-10C_{2D-6}^{p}}
\eea
For the left term of the above inequality, we have
\bea
&&\frac{(4D-p-10)C_{4D-10}^{p}+(3D-p-7)2 C_{3D-7}^{p}-(2D-p-4)C_{2D-4}^{p}}{C_{4D-10}^{p}+2 C_{3D-7}^{p}-C_{2D-4}^{p}}\nn\\
&=&4D-p-10 +\frac{(6-2D) (C_{3D-7}^{p}-C_{2D-4}^{p})}{C_{4D-10}^{p}+2 C_{3D-7}^{p}-C_{2D-4}^{p}}<4D-p-10.
\eea
For the right term of the inequality \eqref{D1-7-factor}, we have
\bea
&&\frac{a_5C_{5D-15}^{p}-a_45C_{4D-12}^{p}+a_310C_{3D-9}^{p}-a_210C_{2D-6}^{p}}{C_{5D-15}^{p}-5C_{4D-12}^{p}+10C_{3D-9}^p-10C_{2D-6}^{p}}\nn\\
&=&a_4+\frac{(D-3)C_{5D-15}^{p}+(3-D)10C_{3D-9}^{p}-(6-2D)10C_{2D-6}^{p}}{C_{5D-15}^{p}-5C_{4D-12}^{p}+10C_{3D-9}^p-10C_{2D-6}^{p}}\nn\\
&=&4D-p-12+\frac{(D-3)(C_{5D-15}^{p}-10C_{3D-9}^{p}+20C_{2D-6}^{p})}{C_{5D-15}^{p}-5C_{4D-12}^{p}+10C_{3D-9}^p-10C_{2D-6}^{p}}\nn\\
&>&4D-p-12+(D-3)/2>4D-p-10.
\eea
In the above proof, it is based on the following results,
\bea
&&\frac{C_{5D-15}^{p}-10C_{3D-9}^{p}+20C_{2D-6}^{p}}{C_{5D-15}^{p}-5C_{4D-12}^{p}+10C_{3D-9}^p-10C_{2D-6}^{p}}>1/2\nn\\
&\Leftrightarrow&
2(C_{5D-15}^{p}-10C_{3D-9}^{p}+20C_{2D-6}^{p})>C_{5D-15}^{p}-5C_{4D-12}^{p}+10C_{3D-9}^p-10C_{2D-6}^{p}\nn\\
&\Leftrightarrow&
C_{5D-15}^{p}+5C_{4D-12}^{p}-30C_{3D-9}^{p}+50C_{2D-6}^{p}>0,
\eea
and
\bea
&&C_{5D-15}^{p}+5C_{4D-12}^{p}-30C_{3D-9}^{p}+50C_{2D-6}^{p}\nn\\
&>&12C_{3D-9}^{p}+5*4C_{3D-9}^{p}-30C_{3D-9}^{p}+50C_{2D-6}^{p}\nn\\
&=&2C_{3D-9}^{p}+50C_{2D-6}^{p}>0,
\eea
where the inequalities \eqref{c5c4c3} are used. So the right term is greater than the left term of the inequality \eqref{D1-7-factor}. The $D_1$ term in the difference \eqref{diff-7} is positive.

Based on the positivity of the four terms in the difference \eqref{diff-7}, we obtain that when $2D-6>p>D-3$,
\bea\label{f-10}
\text{sign}(b_{p})\geqslant \text{sign}(b_{p+1}).
\eea

\subsection{Coefficients of $z^p$, $ p=D-2, D-3$}
In this subsection, we consider the coefficients of $z^p$, $ p=D-2, D-3$, which are
\bea
b_{D-2}&=&a_5 C_{5D-15}^{D-2}r_h^{4D-13}+a'_4 C_{4D-10}^{D-2}r_h^{3D-8}+a_4C_{4D-12}^{D-2}r_h^{3D-10}
\nn\\&&+a'_3 C_{3D-7}^{D-2}r_h^{2D-5}+a_3 C_{3D-9}^{D-2} r_h^{2D-7}+a'_2 C_{2D-4}^{D-2}r_h^{D-2}
\nn\\&&+a_2C_{2D-6}^{D-2}r_h^{D-4}\nn\\
&=& r_h^{-D+2}(-m^5 D_1)(C_{5D-15}^{D-2}-5C_{4D-12}^{D-2}+10C_{3D-9}^{D-2}-10C_{2D-6}^{D-2})\nn\\
&+&r_h^{-D+2}(-4m^5\l_l)(C_{5D-15}^{D-2}+(D-6)C_{4D-12}^{D-2}+(9-2D)C_{3D-9}^{D-2}+(D-4)C_{2D-6}^{D-2})\nn\\
&+&r_h^{-D+4}4(D-3)m^5(C_{4D-10}^{D-2}+2 C_{3D-7}^{D-2}-C_{2D-4}^{D-2})(c_D e- \o)(\frac{(C_{3D-7}^{D-2}+C_{2D-4}^{D-2}) c_D e}{C_{4D-10}^{D-2}+2 C_{3D-7}^{D-2}-C_{2D-4}^{D-2}}-\o)\nn\\
&+&r_h^{-D+4}4(D-3)m^5(\mu^2-\o^2)(2C_{3D-7}^{D-2}-C_{4D-10}^{D-2}-C_{2D-4}^{D-2}),
\eea
\bea
b_{D-3}&=&a_5 C_{5D-15}^{D-3}r_h^{4D-12}+a'_4 C_{4D-10}^{D-3}r_h^{3D-7}+a_4C_{4D-12}^{D-3}r_h^{3D-9}
\nn\\&&+a'_3 C_{3D-7}^{D-3}r_h^{2D-4}+a_3 C_{3D-9}^{D-3} r_h^{2D-6}+a'_2 C_{2D-4}^{D-3}r_h^{D-1}
\nn\\&&+a_2C_{2D-6}^{D-3}r_h^{D-3}+a_1C_{D-3}^{D-3}\nn\\
&=& (-m^4 D_1)(C_{5D-15}^{D-3}-5C_{4D-12}^{D-3}+10C_{3D-9}^{D-3}-10C_{2D-6}^{D-3}+5)\nn\\
&+&(-4m^4\l_l)(C_{5D-15}^{D-3}+(D-6)C_{4D-12}^{D-3}+(9-2D)C_{3D-9}^{D-3}+(D-4)C_{2D-6}^{D-3})\nn\\
&+&r_h^{2}4(D-3)m^4(C_{4D-10}^{D-3}+2 C_{3D-7}^{D-3}-C_{2D-4}^{D-3})(c_D e- \o)(\frac{(C_{3D-7}^{D-3}+C_{2D-4}^{D-3}) c_D e}{C_{4D-10}^{D-3}+2 C_{3D-7}^{D-3}-C_{2D-4}^{D-3}}-\o)\nn\\
&+&r_h^{2}4(D-3)m^4(\mu^2-\o^2)(2C_{3D-7}^{D-3}-C_{4D-10}^{D-3}-C_{2D-4}^{D-3}).
\eea

Now we consider the difference between two normalized coefficients, $\tb_{D-3},\tb_{D-2}$,
\bea\label{diff-8}
\tb_{D-3}-\tb_{D-2}=\frac{ b_{D-3}}{C_{4D-10}^{D-3}+2 C_{3D-7}^{D-3}-C_{2D-4}^{D-3}}
-\frac{r_h b_{D-2}}{C_{4D-10}^{D-2}+2 C_{3D-7}^{D-2}-C_{2D-4}^{D-2}}.
\eea
The difference can be decomposed into four terms and we will analyze term by term.

First, let's see the $(\mu^2-\o^2)$ term in the difference \eqref{diff-8}, which is
\bea
r_h^{2}4(D-3)m^4(\mu^2-\o^2)[\frac{2C_{3D-7}^{D-3}-C_{4D-10}^{D-3}-C_{2D-4}^{D-3}}{C_{4D-10}^{D-3}+2 C_{3D-7}^{D-3}-C_{2D-4}^{D-3}}\nn\\ \label{mu-o-8}
-\frac{2C_{3D-7}^{D-2}-C_{4D-10}^{D-2}-C_{2D-4}^{D-2}}{C_{4D-10}^{D-2}+2 C_{3D-7}^{D-2}-C_{2D-4}^{D-2}}]
\eea
The second line in the above square bracket can be rewritten as
\bea\nn
\frac{2C_{3D-7}^{D-2}-C_{4D-10}^{D-2}-C_{2D-4}^{D-2}}{C_{4D-10}^{D-2}+2 C_{3D-7}^{D-2}-C_{2D-4}^{D-2}}=
\frac{(2D-4)2C_{3D-7}^{D-3}-(3D-7)C_{4D-10}^{D-3}-(D-1)C_{2D-4}^{D-3}}{(3D-7)C_{4D-10}^{D-3}+(2D-4)2 C_{3D-7}^{D-3}-(D-1)C_{2D-4}^{D-3}}.
\eea
A straightforward calculation of the factor in the square bracket in \eqref{mu-o-8} is
\bea\nn
\frac{4(D-3)C_{4D-10}^{D-3}(C_{3D-7}^{D-3}-C_{2D-4}^{D-3})}{(C_{4D-10}^{D-3}+2 C_{3D-7}^{D-3}-C_{2D-4}^{D-3})((3D-7)C_{4D-10}^{D-3}+(2D-4)2 C_{3D-7}^{D-3}-(D-1)C_{2D-4}^{D-3})}
>0.
\eea
So the $(\mu^2-\o^2)$ term in the difference \eqref{diff-8} is positive given the bound state condition.

Secondly, let's see the $(c_D e-\o)$ term in the difference \eqref{diff-8},which is
\bea
r_h^{2}4(D-3)m^4(c_D e- \o)c_D e[\frac{C_{3D-7}^{D-3}+C_{2D-4}^{D-3}}{C_{4D-10}^{D-3}+2 C_{3D-7}^{D-3}-C_{2D-4}^{D-3}}\nn\\ \label{e-o-8}
-\frac{C_{3D-7}^{D-2}+C_{2D-4}^{D-2}}{C_{4D-10}^{D-2}+2 C_{3D-7}^{D-2}-C_{2D-4}^{D-2}}]
\eea
The second line in the above square bracket can be rewritten as
\bea\nn
\frac{C_{3D-7}^{D-2}+C_{2D-4}^{D-2}}{C_{4D-10}^{D-2}+2 C_{3D-7}^{D-2}-C_{2D-4}^{D-2}}=
\frac{(2D-4)C_{3D-7}^{D-3}+(D-1)C_{2D-4}^{D-3}}{(3D-7)C_{4D-10}^{D-3}+(2D-4)2 C_{3D-7}^{D-3}-(D-1)C_{2D-4}^{D-3}}.
\eea
A straightforward calculation of the factor in the square bracket in \eqref{e-o-8} is
\bea\nn
\frac{(D-3)(C_{4D-10}^{D-3}C_{3D-7}^{D-3}+2C_{4D-10}^{D-3}C_{2D-4}^{D-3}+3C_{3D-7}^{D-3}C_{2D-4}^{D-3})}{(C_{4D-10}^{D-3}+2 C_{3D-7}^{D-3}-C_{2D-4}^{D-3})((3D-7)C_{4D-10}^{D-3}+(2D-4)2 C_{3D-7}^{D-3}-(D-1)C_{2D-4}^{D-3})}>0.
\eea
So the $(c_D e-\o)$ term in the difference \eqref{diff-8} is positive given the superradiance condition.

Thirdly, let's see the $\l_l$ term in the difference \eqref{diff-8},which is
\bea
-4m^4\l_l[\frac{C_{5D-15}^{D-3}+(D-6)C_{4D-12}^{D-3}+(9-2D)C_{3D-9}^{D-3}+(D-4)C_{2D-6}^{D-3}}{C_{4D-10}^{D-3}+2 C_{3D-7}^{D-3}-C_{2D-4}^{D-3}}\nn\\
-\frac{C_{5D-15}^{D-2}+(D-6)C_{4D-12}^{D-2}+(9-2D)C_{3D-9}^{D-2}+(D-4)C_{2D-6}^{D-2}}{C_{4D-10}^{D-2}+2 C_{3D-7}^{D-2}-C_{2D-4}^{D-2}}]
\eea
The positivity of the above inequality is equivalent to the negativity of the factor in the square bracket, i.e.
\bea
&&\frac{C_{5D-15}^{D-3}+(D-6)C_{4D-12}^{D-3}+(9-2D)C_{3D-9}^{D-3}+(D-4)C_{2D-6}^{D-3}}{C_{4D-10}^{D-3}+2 C_{3D-7}^{D-3}-C_{2D-4}^{D-3}}\nn\\
&&<\frac{C_{5D-15}^{D-2}+(D-6)C_{4D-12}^{D-2}+(9-2D)C_{3D-9}^{D-2}+(D-4)C_{2D-6}^{D-2}}{C_{4D-10}^{D-2}+2 C_{3D-7}^{D-2}-C_{2D-4}^{D-2}}\nn\\
&\Leftrightarrow&
\frac{C_{4D-10}^{D-2}+2 C_{3D-7}^{D-2}-C_{2D-4}^{D-2}}{C_{4D-10}^{D-3}+2 C_{3D-7}^{D-3}-C_{2D-4}^{D-3}}\nn\\
&&<\frac{C_{5D-15}^{D-2}+(D-6)C_{4D-12}^{D-2}+(9-2D)C_{3D-9}^{D-2}+(D-4)C_{2D-6}^{D-2}}{C_{5D-15}^{D-3}+(D-6)C_{4D-12}^{D-3}+(9-2D)C_{3D-9}^{D-3}+(D-4)C_{2D-6}^{D-3}}
\nn\\
&\Leftrightarrow&
\frac{(3D-7)C_{4D-10}^{D-3}+(2D-4)2 C_{3D-7}^{D-3}-(D-1)C_{2D-4}^{D-3}}{C_{4D-10}^{D-3}+2 C_{3D-7}^{D-3}-C_{2D-4}^{D-3}}\nn\\ \label{l-8-factor}
&&<\frac{a_4C_{5D-15}^{D-3}+a_3(D-6)C_{4D-12}^{D-3}+a_2(9-2D)C_{3D-9}^{D-3}+a_1(D-4)C_{2D-6}^{D-3}}{C_{5D-15}^{D-3}+(D-6)C_{4D-12}^{D-3}+(9-2D)C_{3D-9}^{D-3}+(D-4)C_{2D-6}^{D-3}},
\eea
where $a_i=i(D-3)$.
For the left term of the above inequality, we have
\bea
&&\frac{(3D-7)C_{4D-10}^{D-3}+(2D-4)2 C_{3D-7}^{D-3}-(D-1)C_{2D-4}^{D-3}}{C_{4D-10}^{D-3}+2 C_{3D-7}^{D-3}-C_{2D-4}^{D-3}}\nn\\
&=&3D-7+\frac{(6-2D)( C_{3D-7}^{D-3}-C_{2D-4}^{D-3})}{C_{4D-10}^{D-3}+2 C_{3D-7}^{D-3}-C_{2D-4}^{D-3}}<3D-7.
\eea
For the right term of the inequality \eqref{l-8-factor}, we have
\bea
&&\frac{a_4C_{5D-15}^{D-3}+a_3(D-6)C_{4D-12}^{D-3}+a_2(9-2D)C_{3D-9}^{D-3}+a_1(D-4)C_{2D-6}^{D-3}}{C_{5D-15}^{D-3}+(D-6)C_{4D-12}^{D-3}+(9-2D)C_{3D-9}^{D-3}+(D-4)C_{2D-6}^{D-3}}
\nn\\
&&=3D-7+\nn\\
&&\frac{(D-5)C_{5D-15}^{D-3}-2(D-6)C_{4D-12}^{D-3}+(1-D)(9-2D)C_{3D-9}^{D-3}+(4-2D)(D-4)C_{2D-6}^{D-3}}{C_{5D-15}^{D-3}+(D-6)C_{4D-12}^{D-3}+(9-2D)C_{3D-9}^{D-3}+(D-4)C_{2D-6}^{D-3}}
\nn\\
&&>3D-7.
\eea
In the above, we use the following result
\bea
&&(D-5)C_{5D-15}^{D-3}-2(D-6)C_{4D-12}^{D-3}+(1-D)(9-2D)C_{3D-9}^{D-3}+(4-2D)(D-4)C_{2D-6}^{D-3}\nn\\
&=&(D-5)C_{5D-15}^{D-3}-2(D-6)C_{4D-12}^{D-3}+(D-7)C_{3D-9}^{D-3}+(2D-4)(D-4)(C_{3D-9}^{D-3}-C_{2D-6}^{D-3})\nn\\
&>&C_{5D-15}^{D-3}(D-5-(2D-12)(4/5)^{D-3})+(D-7)C_{3D-9}^{D-3}+(2D-4)(D-4)(C_{3D-9}^{D-3}-C_{2D-6}^{D-3})\nn\\
&>&0.\nn
\eea
So the right term is greater than the left term of the inequality \eqref{l-8-factor}. The $\l_l$ term in the difference \eqref{diff-8} is positive.

Finally, let's see the $D_1$ term in the difference \eqref{diff-8}, which is
\bea
-m^4D_1[\frac{C_{5D-15}^{D-3}-5C_{4D-12}^{D-3}+10C_{3D-9}^{D-3}-10C_{2D-6}^{D-3}+5}{C_{4D-10}^{D-3}+2 C_{3D-7}^{D-3}-C_{2D-4}^{D-3}}\nn\\
-\frac{C_{5D-15}^{D-2}-5C_{4D-12}^{D-2}+10C_{3D-9}^{D-2}-10C_{2D-6}^{D-2}}{C_{4D-10}^{D-2}+2 C_{3D-7}^{D-2}-C_{2D-4}^{D-2}}]
\eea
For $D=7$, one can check directly that the above term is positive. In the next, we discuss $D\geqslant 8$ cases.

The positivity of the $D_1$ term is equivalent to the negativity of the factor in square bracket, i.e.
\bea
&&\frac{C_{5D-15}^{D-3}-5C_{4D-12}^{D-3}+10C_{3D-9}^{D-3}-10C_{2D-6}^{D-3}+5}{C_{4D-10}^{D-3}+2 C_{3D-7}^{D-3}-C_{2D-4}^{D-3}}
<\frac{C_{5D-15}^{D-2}-5C_{4D-12}^{D-2}+10C_{3D-9}^{D-2}-10C_{2D-6}^{D-2}}{C_{4D-10}^{D-2}+2 C_{3D-7}^{D-2}-C_{2D-4}^{D-2}}\nn\\
&\Leftrightarrow&
\frac{C_{4D-10}^{D-2}+2 C_{3D-7}^{D-2}-C_{2D-4}^{D-2}}{C_{4D-10}^{D-3}+2 C_{3D-7}^{D-3}-C_{2D-4}^{D-3}}
<\frac{C_{5D-15}^{D-2}-5C_{4D-12}^{D-2}+10C_{3D-9}^{D-2}-10C_{2D-6}^{D-2}}{C_{5D-15}^{D-3}-5C_{4D-12}^{D-3}+10C_{3D-9}^{D-3}-10C_{2D-6}^{D-3}+5}\nn\\
&\Leftrightarrow&
\frac{(3D-7)C_{4D-10}^{D-3}+(2D-4)2 C_{3D-7}^{D-3}-(D-1)C_{2D-4}^{D-3}}{C_{4D-10}^{D-3}+2 C_{3D-7}^{D-3}-C_{2D-4}^{D-3}}\nn\\ \label{D1-8-factor}
&&<\frac{(4D-12)C_{5D-15}^{D-3}-(3D-9) 5C_{4D-12}^{D-3}+(2D-6)10C_{3D-9}^{D-3}-(D-3)10C_{2D-6}^{D-3}}{C_{5D-15}^{D-3}-5C_{4D-12}^{D-3}+10C_{3D-9}^{D-3}-10C_{2D-6}^{D-3}+5}.
\eea
For the left term of the above inequality, we have
\bea
&&\frac{(3D-7)C_{4D-10}^{D-3}+(2D-4)2 C_{3D-7}^{D-3}-(D-1)C_{2D-4}^{D-3}}{C_{4D-10}^{D-3}+2 C_{3D-7}^{D-3}-C_{2D-4}^{D-3}}\nn\\
&=&3D-7+\frac{(6-2D)( C_{3D-7}^{D-3}-C_{2D-4}^{D-3})}{C_{4D-10}^{D-3}+2 C_{3D-7}^{D-3}-C_{2D-4}^{D-3}}<3D-7.
\eea
For the right term of the inequality \eqref{D1-8-factor}, we have
\bea
&&\frac{(4D-12)C_{5D-15}^{D-3}-(3D-9) 5C_{4D-12}^{D-3}+(2D-6)10C_{3D-9}^{D-3}-(D-3)10C_{2D-6}^{D-3}}{C_{5D-15}^{D-3}-5C_{4D-12}^{D-3}+10C_{3D-9}^{D-3}-10C_{2D-6}^{D-3}+5}\nn\\
&=&3D-7\nn\\
&+&\frac{(D-5)C_{5D-15}^{D-3}+10C_{4D-12}^{D-3}-(D-1)10C_{3D-9}^{D-3}+(2D-4)10C_{2D-6}^{D-3}-5(3D-7)}{C_{5D-15}^{D-3}-5C_{4D-12}^{D-3}+10C_{3D-9}^{D-3}-10C_{2D-6}^{D-3}+5}\nn\\
&>&3D-7.
\eea
In the last line of the above equation, we use the positivity of the following term
\bea
\frac{(D-5)C_{5D-15}^{D-3}+10C_{4D-12}^{D-3}-(D-1)10C_{3D-9}^{D-3}+(2D-4)10C_{2D-6}^{D-3}-5(3D-7)}{C_{5D-15}^{D-3}-5C_{4D-12}^{D-3}+10C_{3D-9}^{D-3}-10C_{2D-6}^{D-3}+5}.
\nn\\
\eea
The positivity of the denominator of the above term can be checked directly for $D=8,9,10$ and for $D>10$, since $C_{5D-15}^{D-3}/C_{4D-12}^{D-3}>(5/4)^{D-3}>5$, the denominator is positive. For the numerator of the above term, we have
\bea
&&[(D-5)C_{5D-15}^{D-3}+10C_{4D-12}^{D-3}-(D-1)10C_{3D-9}^{D-3}]+[(2D-4)10C_{2D-6}^{D-3}-5(3D-7)]\nn\\
&>&[(D-5)(\frac{5}{3})^{D-3}+10(4/3)^{D-3}-10(D-1)]C_{3D-9}^{D-3}+[(2D-4)10C_{2D-6}^{D-3}-5(3D-7)].\nn
\eea
The factor in the first square bracket is positive when $D>7$ and the term in the second square bracket is obviously positive, then the numerator is also positive.

So the right term is greater than the left term in the inequality \eqref{D1-8-factor} and the $D_1$ term in the difference \eqref{diff-8} is positive.

Based on the positivity of the four terms in the difference \eqref{diff-8}, we obtain that
\bea\label{f-11}
\text{sign}(b_{D-3})\geqslant \text{sign}(b_{D-2}).
\eea

\subsection{Coefficients of $z^p$, $ D-3> p >0 $}
When $ D-3> p >0 $, the coefficient of $z^p$ is
\bea
b_{p}&=&a_5 C_{5D-15}^{p}r_h^{5D-15-p}+a'_4 C_{4D-10}^{p}r_h^{4D-10-p}+a_4C_{4D-12}^{p}r_h^{4D-12-p}
\nn\\&&+a'_3 C_{3D-7}^{p}r_h^{3D-7-p}+a_3 C_{3D-9}^{p} r_h^{3D-9-p}+a'_2 C_{2D-4}^{p}r_h^{2D-4-p}
\nn\\&&+a_2C_{2D-6}^{p}r_h^{2D-6-p}+a_1C_{D-3}^{p}r_h^{D-3-p}\nn\\
&=& r_h^{-p}(-m^5 D_1)(C_{5D-15}^{p}-5C_{4D-12}^{p}+10C_{3D-9}^p-10C_{2D-6}^{p}+5C_{D-3}^{p})\nn\\
&+&r_h^{-p}(-4m^5\l_l)(C_{5D-15}^{p}+(D-6)C_{4D-12}^{p}+(9-2D)C_{3D-9}^p+(D-4)C_{2D-6}^{p})\nn\\
&+&r_h^{-p+2}4(D-3)m^5(C_{4D-10}^{p}+2 C_{3D-7}^{p}-C_{2D-4}^{p})(c_D e- \o)(\frac{(C_{3D-7}^{p}+C_{2D-4}^{p}) c_D e}{C_{4D-10}^{p}+2 C_{3D-7}^{p}-C_{2D-4}^{p}}-\o)\nn\\
&+&r_h^{-p+2}4(D-3)m^5(\mu^2-\o^2)(2C_{3D-7}^{p}-C_{4D-10}^{p}-C_{2D-4}^{p}).
\eea
Now we consider the difference between two normalized coefficients, $\tb_{p},\tb_{p+1}$,
\bea\label{diff-9}
\tilde{b}_p-\tilde{b}_{p+1}=\frac{r_h^p b_p}{C_{4D-10}^{p}+2 C_{3D-7}^{p}-C_{2D-4}^{p}}-\frac{r_h^{p+1} b_{p+1}}{C_{4D-10}^{p+1}+2 C_{3D-7}^{p+1}-C_{2D-4}^{p+1}}
\eea
The difference can be decomposed into four terms and we will analyze term by term.

First, let's see the $(\mu^2-\o^2)$ term in the difference \eqref{diff-9}, which is
\bea
r_h^{2}4(D-3)m^5(\mu^2-\o^2)[\frac{2C_{3D-7}^{p}-C_{4D-10}^{p}-C_{2D-4}^{p}}{C_{4D-10}^{p}+2 C_{3D-7}^{p}-C_{2D-4}^{p}}\nn\\
-\frac{2C_{3D-7}^{p+1}-C_{4D-10}^{p+1}-C_{2D-4}^{p+1}}{C_{4D-10}^{p+1}+2 C_{3D-7}^{p+1}-C_{2D-4}^{p+1}}]
\eea
The positivity of the above term is equivalent to the positivity of the factor in the square bracket. After a straightforward calculation of this factor, we obtain
\bea
&&\frac{2C_{3D-7}^{p}-C_{4D-10}^{p}-C_{2D-4}^{p}}{C_{4D-10}^{p}+2 C_{3D-7}^{p}-C_{2D-4}^{p}}
-\frac{2C_{3D-7}^{p+1}-C_{4D-10}^{p+1}-C_{2D-4}^{p+1}}{C_{4D-10}^{p+1}+2 C_{3D-7}^{p+1}-C_{2D-4}^{p+1}}\nn\\
&=&\frac{4(D-3)C_{4D-10}^{p}(C_{3D-7}^{p}-C_{2D-4}^{p})}{(C_{4D-10}^{p}+2 C_{3D-7}^{p}-C_{2D-4}^{p})(a'_4C_{4D-10}^{p}+a'_3 2 C_{3D-7}^{p}-a'_2 C_{2D-4}^{p})}
>0,
\eea
where $a'_4=4D-p-10, a'_3=3D-p-7, a'_2=2D-p-4$. So the $(\mu^2-\o^2)$ term in the difference \eqref{diff-9} is positive given the bound state condition.

Secondly, let's see the $(c_D e-\o)$ term in the difference \eqref{diff-9}, which is
\bea
r_h^{2}4(D-3)m^5(c_D e- \o)c_D e[\frac{C_{3D-7}^{p}+C_{2D-4}^{p}}{C_{4D-10}^{p}+2 C_{3D-7}^{p}-C_{2D-4}^{p}}\nn\\
-\frac{C_{3D-7}^{p+1}+C_{2D-4}^{p+1}}{C_{4D-10}^{p+1}+2 C_{3D-7}^{p+1}-C_{2D-4}^{p+1}}]
\eea
The positivity of the above term is equivalent to the positivity of the factor in the square bracket. After a straightforward calculation of this factor, we obtain
\bea
&&\frac{C_{3D-7}^{p}+C_{2D-4}^{p}}{C_{4D-10}^{p}+2 C_{3D-7}^{p}-C_{2D-4}^{p}}
-\frac{C_{3D-7}^{p+1}+C_{2D-4}^{p+1}}{C_{4D-10}^{p+1}+2 C_{3D-7}^{p+1}-C_{2D-4}^{p+1}}\nn\\
&=&\frac{(D-3)(C_{4D-10}^{p}C_{3D-7}^{p}+2C_{4D-10}^{p}C_{2D-4}^{p}+3C_{3D-7}^{p}C_{2D-4}^{p})}{(C_{4D-10}^{p}+2 C_{3D-7}^{p}-C_{2D-4}^{p})(a'_4C_{4D-10}^{p}+a'_3 2 C_{3D-7}^{p}-a'_2 C_{2D-4}^{p})}>0.
\eea
So the $(c_D e-\o)$ term in the difference \eqref{diff-9} is positive given the superradiance condition.

Thirdly, let's see the $\l_l$ term in the difference \eqref{diff-9}, which is
\bea
-4m^5\l_l[\frac{C_{5D-15}^{p}+(D-6)C_{4D-12}^{p}+(9-2D)C_{3D-9}^p+(D-4)C_{2D-6}^{p}}{C_{4D-10}^{p}+2 C_{3D-7}^{p}-C_{2D-4}^{p}}\nn\\
-\frac{C_{5D-15}^{p+1}+(D-6)C_{4D-12}^{p+1}+(9-2D)C_{3D-9}^{p+1}+(D-4)C_{2D-6}^{p+1}}{C_{4D-10}^{p+1}+2 C_{3D-7}^{p+1}-C_{2D-4}^{p+1}}]
\eea
The positivity of the above term is equivalent to the negativity of the factor in the square bracket, i.e.
\bea
&&\frac{C_{5D-15}^{p}+(D-6)C_{4D-12}^{p}+(9-2D)C_{3D-9}^p+(D-4)C_{2D-6}^{p}}{C_{4D-10}^{p}+2 C_{3D-7}^{p}-C_{2D-4}^{p}}\nn\\
&&<\frac{C_{5D-15}^{p+1}+(D-6)C_{4D-12}^{p+1}+(9-2D)C_{3D-9}^{p+1}+(D-4)C_{2D-6}^{p+1}}{C_{4D-10}^{p+1}+2 C_{3D-7}^{p+1}-C_{2D-4}^{p+1}}\nn\\
&\Leftrightarrow&
\frac{C_{4D-10}^{p+1}+2 C_{3D-7}^{p+1}-C_{2D-4}^{p+1}}{C_{4D-10}^{p}+2 C_{3D-7}^{p}-C_{2D-4}^{p}}\nn\\
&&<\frac{C_{5D-15}^{p+1}+(D-6)C_{4D-12}^{p+1}+(9-2D)C_{3D-9}^{p+1}+(D-4)C_{2D-6}^{p+1}}{C_{5D-15}^{p}+(D-6)C_{4D-12}^{p}+(9-2D)C_{3D-9}^p+(D-4)C_{2D-6}^{p}}\nn\\
&\Leftrightarrow&
\frac{(4D-p-10)C_{4D-10}^{p}+(3D-p-7)2 C_{3D-7}^{p}-(2D-p-4)C_{2D-4}^{p}}{C_{4D-10}^{p}+2 C_{3D-7}^{p}-C_{2D-4}^{p}}\nn\\ \label{l-9-factor}
&&<\frac{a_5C_{5D-15}^{p}+a_4(D-6)C_{4D-12}^{p}+a_3(9-2D)C_{3D-9}^{p}+a_2(D-4)C_{2D-6}^{p}}{C_{5D-15}^{p}+(D-6)C_{4D-12}^{p}+(9-2D)C_{3D-9}^p+(D-4)C_{2D-6}^{p}}.
\eea
In the above inequality, $a_i=(D-3)i-p$. For the left term of the above inequality, we have
\bea
&&\frac{(4D-p-10)C_{4D-10}^{p}+(3D-p-7)2 C_{3D-7}^{p}-(2D-p-4)C_{2D-4}^{p}}{C_{4D-10}^{p}+2 C_{3D-7}^{p}-C_{2D-4}^{p}}\nn\\
&=&4D-p-10+\frac{(6-2D)( C_{3D-7}^{p}-C_{2D-4}^{p})}{C_{4D-10}^{p}+2 C_{3D-7}^{p}-C_{2D-4}^{p}}<4D-p-10.
\eea
For the right term of the inequality \eqref{l-9-factor}, we have
\bea
&&\frac{a_5C_{5D-15}^{p}+a_4(D-6)C_{4D-12}^{p}+a_3(9-2D)C_{3D-9}^{p}+a_2(D-4)C_{2D-6}^{p}}{C_{5D-15}^{p}+(D-6)C_{4D-12}^{p}+(9-2D)C_{3D-9}^p+(D-4)C_{2D-6}^{p}}\nn\\
&=&a_4+\frac{(D-3)C_{5D-15}^{p}+(3-D)(9-2D)C_{3D-9}^{p}+2(3-D)(D-4)C_{2D-6}^{p}}{C_{5D-15}^{p}+(D-6)C_{4D-12}^{p}+(9-2D)C_{3D-9}^p+(D-4)C_{2D-6}^{p}}\nn\\
&=&4D-p-10\nn\\
&+&\frac{(D-5)C_{5D-15}^{p}-2(D-6)C_{4D-12}^{p}+(D-1)(2D-9)C_{3D-9}^{p}-(2D-4)(D-4)C_{2D-6}^{p}}{C_{5D-15}^{p}+(D-6)C_{4D-12}^{p}+(9-2D)C_{3D-9}^p+(D-4)C_{2D-6}^{p}}\nn\\
&>&4D-p-10.
\eea
In the last line of the above inequality,  we need the following result
\bea
&&(D-5)C_{5D-15}^{p}-2(D-6)C_{4D-12}^{p}+(D-1)(2D-9)C_{3D-9}^{p}-(2D-4)(D-4)C_{2D-6}^{p}>0\nn\\
&\Leftrightarrow&
(D-5)C_{5D-15}^{p}-2(D-6)C_{4D-12}^{p}+(D-7)C_{3D-9}^{p}+(2D-4)(D-4)(C_{3D-9}^{p}-C_{2D-6}^{p})>0\nn\\
&\Leftrightarrow&
(D-5)(C_{5D-15}^{p}-C_{4D-12}^{p})+(D-7)(C_{3D-9}^{p}-C_{4D-12}^{p})+(2D-4)(D-4)(C_{3D-9}^{p}-C_{2D-6}^{p})>0\nn\\
&\Leftrightarrow&
(D-5)(C_{5D-15}^{p}+C_{3D-9}^{p}-2C_{4D-12}^{p})\nn\\ \label{l-9-factor-1}
&&+2(C_{4D-12}^{p}-C_{3D-9}^{p})+(2D-4)(D-4)(C_{3D-9}^{p}-C_{2D-6}^{p})>0
\eea
It is easy to see that if the first line of the above inequality \eqref{l-9-factor-1} is non-negative and then the above inequality holds. The first line can be rewritten as
\bea
(D-5)(C_{5D-15}^{p}+C_{3D-9}^{p}-2C_{4D-12}^{p})=(D-5)C_{4D-12}^{p}(\frac{C_{5D-15}^{p}}{C_{4D-12}^{p}}+\frac{C_{3D-9}^{p}}{C_{4D-12}^{p}}-2).\nn\\
\eea
It is easy to check that when $p>3$
\bea
\frac{C_{5D-15}^{p}}{C_{4D-12}^{p}}>(5/4)^p>2.
\eea
And when $p=1$,
\bea
\frac{C_{5D-15}^{1}}{C_{4D-12}^{1}}+\frac{C_{3D-9}^{1}}{C_{4D-12}^{1}}-2=0.
\eea
When $p=2$,
\bea
\frac{C_{5D-15}^{2}}{C_{4D-12}^{2}}+\frac{C_{3D-9}^{2}}{C_{4D-12}^{2}}-2
=\frac{5}{4}\cdot\frac{5D-16}{4D-13}+\frac{3}{4}\cdot\frac{3D-10}{4D-13}-2.
\eea
It is easy to check the above expression is positive when $D>6$. Similar result holds for $p=3$. Then the  inequality \eqref{l-9-factor-1} holds.

And the right term is greater than the left term of the inequality \eqref{l-9-factor}. The $\l_l$ term in the difference \eqref{diff-9} is positive.

Finally, let's see the $D_1$ term in the difference \eqref{diff-9}, which is
\bea
-m^5D_1[\frac{C_{5D-15}^{p}-5C_{4D-12}^{p}+10C_{3D-9}^p-10C_{2D-6}^{p}+5C_{D-3}^{p}}{C_{4D-10}^{p}+2 C_{3D-7}^{p}-C_{2D-4}^{p}}\nn\\
-\frac{C_{5D-15}^{p+1}-5C_{4D-12}^{p+1}+10C_{3D-9}^{p+1}-10C_{2D-6}^{p+1}+5C_{D-3}^{p+1}}{C_{4D-10}^{p+1}+2 C_{3D-7}^{p+1}-C_{2D-4}^{p+1}}].
\eea
One can check that when $1\leqslant p\leqslant 4$,
\bea
C_{5D-15}^{p}-5C_{4D-12}^{p}+10C_{3D-9}^p-10C_{2D-6}^{p}+5C_{D-3}^{p}=0.
\eea
and when $p\geqslant5$ and $D>6$
\bea
C_{5D-15}^{p}-5C_{4D-12}^{p}+10C_{3D-9}^p-10C_{2D-6}^{p}+5C_{D-3}^{p}>0.
\eea
For the above inequality, one can check it directly when $p=5,6,7$. When $p\geqslant 8$, $C_{5D-15}^{p}/C_{4D-12}^{p}>(5/4)^8>5$ and the above inequality holds.

Then the $D_1$ term  is non-negative for $1\leqslant p\leqslant 4$. In the next, we only discuss the $p>4$ cases.
The positivity of the $D_1$ term is equivalent to the negativity of the factor in the square bracket, i.e.
\bea
&&\frac{C_{5D-15}^{p}-5C_{4D-12}^{p}+10C_{3D-9}^p-10C_{2D-6}^{p}+5C_{D-3}^{p}}{C_{4D-10}^{p}+2 C_{3D-7}^{p}-C_{2D-4}^{p}}\nn\\
&&<\frac{C_{5D-15}^{p+1}-5C_{4D-12}^{p+1}+10C_{3D-9}^{p+1}-10C_{2D-6}^{p+1}+5C_{D-3}^{p+1}}{C_{4D-10}^{p+1}+2 C_{3D-7}^{p+1}-C_{2D-4}^{p+1}}\nn\\
&\Leftrightarrow&
\frac{C_{4D-10}^{p+1}+2 C_{3D-7}^{p+1}-C_{2D-4}^{p+1}}{C_{4D-10}^{p}+2 C_{3D-7}^{p}-C_{2D-4}^{p}}\nn\\
&&<\frac{C_{5D-15}^{p+1}-5C_{4D-12}^{p+1}+10C_{3D-9}^{p+1}-10C_{2D-6}^{p+1}+5C_{D-3}^{p+1}}{C_{5D-15}^{p}-5C_{4D-12}^{p}+10C_{3D-9}^p-10C_{2D-6}^{p}+5C_{D-3}^{p}}\nn\\
&\Leftrightarrow&
\frac{(4D-p-10)C_{4D-10}^{p}+(3D-p-7)2 C_{3D-7}^{p}-(2D-p-4)C_{2D-4}^{p}}{C_{4D-10}^{p}+2 C_{3D-7}^{p}-C_{2D-4}^{p}}\nn\\ \label{D1-9-factor}
&&<\frac{a_5C_{5D-15}^{p}-a_4 5C_{4D-12}^{p}+a_3 10C_{3D-9}^{p}-a_2 10C_{2D-6}^{p}+a_1 5C_{D-3}^{p}}{C_{5D-15}^{p}-5C_{4D-12}^{p}+10C_{3D-9}^p-10C_{2D-6}^{p}+5C_{D-3}^{p}},
\eea
where $a_i=(D-3)i-p$.

For the left term of the above inequality, we have
\bea
&&\frac{(4D-p-10)C_{4D-10}^{p}+(3D-p-7)2 C_{3D-7}^{p}-(2D-p-4)C_{2D-4}^{p}}{C_{4D-10}^{p}+2 C_{3D-7}^{p}-C_{2D-4}^{p}}\nn\\
&=&4D-p-10+\frac{(6-2D)( C_{3D-7}^{p}-C_{2D-4}^{p})}{C_{4D-10}^{p}+2 C_{3D-7}^{p}-C_{2D-4}^{p}}<4D-p-10.
\eea
For the right term of the inequality \eqref{D1-9-factor}, we have
\bea
&&\frac{a_5C_{5D-15}^{p}-a_4 5C_{4D-12}^{p}+a_3 10C_{3D-9}^{p}-a_2 10C_{2D-6}^{p}+a_1 5C_{D-3}^{p}}{C_{5D-15}^{p}-5C_{4D-12}^{p}+10C_{3D-9}^p-10C_{2D-6}^{p}+5C_{D-3}^{p}}\nn\\
&=&4D-p-10\nn\\
&+&\frac{(D-5)C_{5D-15}^{p}+10C_{4D-12}^{p}-10(D-1)C_{3D-9}^{p}+20(D-2)C_{2D-6}^{p}-5(3D-7)C_{D-3}^{p}}{C_{5D-15}^{p}-5C_{4D-12}^{p}+10C_{3D-9}^p-10C_{2D-6}^{p}+5C_{D-3}^{p}}\nn\\
&>& 4D-p-10.
\eea
In the last line of the above equation, we use the result that when $p>4$
\bea
(D-5)C_{5D-15}^{p}+10C_{4D-12}^{p}-10(D-1)C_{3D-9}^{p}+20(D-2)C_{2D-6}^{p}-5(3D-7)C_{D-3}^{p}
> 0.\nn\\\label{D1-9-factor-1}
\eea
The above inequality can be shown as follows
\bea
&&(D-5)C_{5D-15}^{p}+10C_{4D-12}^{p}-10(D-1)C_{3D-9}^{p}+20(D-2)C_{2D-6}^{p}-5(3D-7)C_{D-3}^{p}\nn\\
&>&C_{3D-9}^{p}[(D-5)(5/3)^p+10(4/3)^p-10(D-1)]+[20(D-2)C_{2D-6}^{p}-5(3D-7)C_{D-3}^{p}]\nn
\eea
When $p\geqslant 5$ and $D>6$, it is easy to check the term in the first square bracket is positive. The term in the second square bracket is obviously positive. Thus the
inequality \eqref{D1-9-factor-1} holds. So the $D_1$ term in the difference \eqref{diff-9} is positive.

Based on the positivity of the four terms in the difference \eqref{diff-9}, we obtain that
\bea\label{f-12}
\text{sign}(b_{p})\geqslant \text{sign}(b_{p+1}).
\eea

\section{Summary}
\begin{figure}[htbp]
\centering
\includegraphics[scale=0.4]{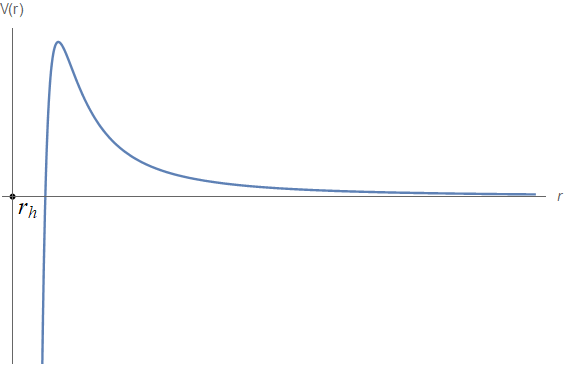}
\caption{A typical shape of the effective potential $V(r)$.}
\label{RN}
\end{figure}

In this work, superradiant stability of  D-dimensional ($D\geq 7$) extremal RN black hole  under charged massive scalar perturbation is studied analytically. Based on
the asymptotic analysis of the effective potential $V(r)$ experienced by the scalar perturbation, we know there is one maximum for the effective potential outside the black hole horizon. Then we derive the numerator $E(z)$ of the derivative of the effective potential, which is a polynomial of $z=r-r_h$ with real coefficients. In Section \eqref{root}, we show in \eqref{f-0}, \eqref{f-1},\eqref{f-2},\eqref{f-3} that
\bea
b_0>0,~~b_p<0~(3D-7<p<5D-15).
\eea
According the results in \eqref{f-4}, \eqref{f-5},\eqref{f-6},\eqref{f-7},\eqref{f-8},\eqref{f-9},\eqref{f-10}, \eqref{f-11},\eqref{f-12}, we obtain
\bea
\text{sign}(b_{p})\geqslant \text{sign}(b_{p+1}),~(0<p<3D-7).
\eea
So the sign change in the following sequence of the real coefficients of $E(z)$,
\bea
(b_{5D-15},b_{5D-16},...,b_{p+1},b_p,..,b_1,b_0),
\eea
is always 1.  Then according to Descartes' rule of signs, we know there is at most 1 positive root for the equation $E(z)=0$. Thus there is only one extreme for the effective potential outside the horizon, which is a maximum and there is no potential well outside the horizon for the superradiance modes. A typical shape of the effective potential outside the horizon is shown in Fig.\eqref{RN}.
The two conditions for the possible superradiant instability can not be satisfied simultaneously, so there is no black hole bomb for D-dimensional extremal RN black hole under charged massive scalar perturbation.

Our result provides a complementary analytical proof for the previous numerical work \cite{Konoplya:2008au,Konoplya:2013sba}. This new method seems to be efficient to analyze the superradiant stability of higher dimensional black holes. We have already applied it to study the superradiant stability of 5-dimensional non-extremal RN black hole under charged massive scalar perturbation in \cite{Huang:2021jaz}. As a step further, it is interesting to apply it to study other higher dimensional non-extremal RN black hole cases and even the D-dimensional ($D\geqslant 5$) non-extremal RN black hole cases. It may also be interesting to study higher dimensional rotating black hole cases.

\acknowledgments
This work is partially supported by Guangdong Major Project of Basic and Applied Basic Research (No. 2020B0301030008), Science and Technology Program of Guangzhou (No. 2019050001) and Natural Science Foundation of Guangdong Province (No. 2020A1515010388, No. 2020A1515010794).

\end{document}